\documentclass[fleqn,10pt]{wlscirep}
\usepackage[utf8]{inputenc}
\usepackage[T1]{fontenc}
\usepackage{multirow}
\usepackage{romannum}
\usepackage{upgreek}
\usepackage{bm}
\usepackage{amssymb}
\usepackage{soul}
\usepackage{xcolor}
\setstcolor{red}
\title{Quantitative predictions of alpha-charmonium correlation functions in high-energy collisions}

\author[1,2,*]{Faisal Etminan}
\affil[1]{Department of Physics, Faculty of Sciences, University of Birjand, P.O.Box 97175-615, Birjand, Iran}
\affil[2]{Interdisciplinary Theoretical and Mathematical Sciences Program (iTHEMS), RIKEN, Wako 351-0198, Japan}

\affil[*]{fetminan@birjand.ac.ir}

\begin{abstract}
	Two-body  $ ^{4}\textrm{He}\left(\alpha\right)$-charmonium $ \left(c\bar{c}\right) $ potentials in the single-folding potential (SFP) approach are built by using a first principles HAL QCD low-energy $ NJ/\psi$ and $ N\eta_{c} $ interactions.
The $N\textrm{-}c\bar{c}$ potentials are observed to exhibit an attractive nature across all distances, accompanied by a characteristic long-range tail. 
It is found that the  $ \alpha\textrm{-}J/\psi $ system appears to be loosely bound 
with the central binding energy in the range of 0.1-0.6   MeV, while for spin-$ 1/2 $ $\alpha\textrm{-}\eta_{c}$, 
no bound or resonance state (with respect to the $ \alpha \textrm{-} c\bar{c} $ threshold) was found.
The $ \alpha\textrm{-}c\bar{c} $ correlation function 
in high-energy collisions is examined to explore the $ N\textrm{-}c\bar{c} $ interaction. 
The analysis revealed that variations in spin-dependent $\alpha\textrm{-}c\bar{c}$ interactions— spin-$3/2$ $\alpha\textrm{-}J/\psi$, spin-$1/2$ $\alpha\textrm{-}J/\psi$, spin-$1/2$ $\alpha\textrm{-}\eta_c$, and the spin-averaged $\alpha\textrm{-}J/\psi$—produce noticeable differences in the $\alpha\textrm{-}c\bar{c}$ correlation function, especially when the source size is around $ 3 $ fm.
It is found that different results are produced by the Lednicky-Lyuboshits formula at small source sizes. 
This indicates that a relatively long-range interaction exists for the $ \alpha\textrm{-}c\bar{c} $ system.
Furthermore, a comparison has been conducted between two density functions of $ ^{4}\textrm{He}$—the central depression (CD) and the simple single Gaussian (SG) density—both of which share an identical rms radius of 1.56 fm. Although the $\alpha\textrm{-}J/\psi$ binding energies for the two models are nearly indistinguishable, their corresponding correlation functions demonstrate markedly different behaviors. This divergence could yield valuable insights into the nuclear matter distribution function of the alpha particle. 
\end{abstract}
\begin{document}

\flushbottom
\maketitle
%
%
\thispagestyle{empty}


\section{Introduction} \label{sec:intro}
Despite extensive research, a proper microscopic mechanism for the formation of light (anti)nuclei ($ A \le 5 $) in high-energy collisions has not yet been established in nuclear physics~\cite{ alice2025reveal}. 
It is not yet fully understood how nuclei—bound by only a few MeV—can form in environments where temperatures exceed 100 MeV, despite extensive research addressing this issue~\cite{andronic2018decoding, alice2025reveal}.

In high-energy ion collisions, particle production is closely related to the confinement of color charge within color-neutral hadrons. These collisions produce a quark–gluon plasma (QGP), which, as it cools and evolves, leads to the formation of hadrons and light nuclei~\cite{andronic2018decoding}. The key questions concern how these loosely-bound nuclei are formed and how they survive through the hadronic phase after the QGP hadronizes.
The yields of light nuclei have been measured at the Relativistic Heavy Ion Collider (RHIC) in Au-Au collisions, at the Large Hadron Collider (LHC) for pp collisions, and additionally for $ pp $-Pb and Pb-Pb collisions~\cite{PhysRevC.93.024917,2020135043, PhysRevLett.134.162301}.
Current understanding suggests that nuclei can be produced either through direct emission as multi-quark states following a collision—similar to other hadrons such as protons or pions—or via a secondary fusion mechanism of nucleons facilitated by mesons~\cite{alice2025reveal}.

Theoretical approaches generally fall into three categories: (i) statistical hadronization models~\cite{andronic2018decoding, PhysRevC.106.044908}, which propose that hadrons and nuclei are produced directly from a thermal and chemically equilibrated source, with their abundances determined by factors such as particle mass, temperature, volume, and quantum number conservation; (ii) coalescence models~\cite{kachelriess2020alternative, PhysRevC.99.044913, mahlein2023realistic}, which posit that nucleons first form independently, then fuse into nuclei through final-state interactions when they are in close proximity in phase space. However, these models do not explicitly incorporate the kinematic conditions, such as energy-momentum conservation, that govern the formation of nuclei. And (iii) kinetic approaches (transport models) simulate the space-time evolution of the collision by solving kinetic equations for hadrons (and sometimes partons), including collision terms, mean fields, and production/dissociation reactions for light nuclei (e.g., deuterons, tritons, helium-3, alpha)~\cite{DANIELEWICZ1991712, PhysRevC.46.2002,PhysRevC.108.L031601, sun2024unveiling, wang2025alphaclusteringwarmdense}. These models can yield event-by-event predictions for yields, spectra, and flow, and are capable of capturing non-equilibrium effects and the time evolution of cluster abundances through the collision.

In heavy-ion experiments, additional and alternative information concerning the elementary hyperon (Y)-nucleon (N) interaction, as well as the lifetimes and binding energies of light hypernuclei, is obtained~\cite{AbdallahPLB2022}. 
In nature, there are cases that are clean, whereas few-body systems involving nucleons are believed to enhance the binding of two-body bound or resonance states~\cite{Garcilazo2019}. 
Consequently, the knowledge of few-body systems is expected to contribute to understanding two-body bound states or resonances within the strange sector. Additionally, while the effect of three-body forces might be insignificant for $A=3$, it is considered to be potentially substantial in bound systems with 
$ A=4\textrm{-}7 $.

The study of $ \textrm{YN} $ interactions through classical scattering experiments is made challenging by the necessity for a stable target and a precisely controlled beam. For interactions involving heavy hadrons containing charm and bottom quarks, which are relevant to the investigation of exotic hadrons, scattering experiments are considered nearly impossible~\cite{hyodo2025femtoscopyexotichadronsnuclei}. 
Even in channels where scattering experiments are feasible, the precision of low-energy data is often limited due to the instability of the beam particles.
It is hoped that femtoscopic techniques will offer a unique opportunity to investigate 
$ \textrm{YN} $ interactions by measuring two-particle correlations~\cite{cho2017exotic} and to provide insights into the space-time geometry of the particle-emitting sources.  Additionally, this technique has higher precision in low-energy correlations than traditional scattering experiments

The correlation function in the strange sector like $ \Lambda p $~\cite{hu2023}, $ \Lambda \Lambda $~\cite{2019134822}, $ N\Xi $~\cite{alice2020unveiling}, $ N\Omega $~\cite{Adam2019, alice2020unveiling}, has provided insight into $ \textrm{YN} $ two-body interactions.
Additionally, correlation functions have been measured in the charm sector,
including the $ D^{-}p $~\cite{PhysRevD.106.052010}, $ D\pi $ and $ DK $ pairs~\cite{PhysRevD.110.032004}.
Access to these channels through traditional scattering experiments is nearly impossible, but valuable information regarding hadron interactions has been obtained through these measurements, complemented by theoretical investigations~\cite{BRODSKY1997125, WU2025, Liang-ZhenPRD2025, liu2025charmoniumnucleonfemto}. 
Specifically, in a series of studies~\cite{krein2020femtoscopy, Krein2022EPJ, krein2023femtoscopy}, it was proposed by Krein et al. that femtoscopic measurements of $J/\psi$-proton correlation functions in high-energy hadron collisions could be utilized to extract information about the low-energy $J/\psi$-nucleon interaction and to investigate the origin of the proton’s mass.

As a subsequent step in femtoscopic analyses, the correlation functions between hadrons and deuterons are considered promising~\cite{PhysRevX.14.031051,GarridoPRC2024, MROWCZYNSKI2025139413} and have been explored theoretically, such as 
$ pd $~\cite{bazak2020production, PhysRevC.108.064002}, $ {K}^{-}d $~\cite{PhysRevX.14.031051, mrowczynski2019hadron},
$ \Lambda d $~\cite{Haidenbauerprc, GarridoPRC2024}, $ \Xi d $~\cite{Ogata2021},
and $ \Omega NN $~\cite{zhang2021production}.
Recently, the momentum correlations involving 
$ \Lambda- $ and $ \Xi-$triton~\cite{EtminanJphysG_2025}, 
$ \Lambda \alpha $~\cite{jinno2024femtoscopic},
$ \Xi \alpha $~\cite{kamiya2024}, $ \Omega \alpha $~\cite{etminan2024omegaAlpha} and 
$ \phi \alpha $~\cite{ETMINAN2025PLB139564} have been investigated theoretically to provide insights into the interactions between hyperons and nucleons.

In addition to recent advances in both theoretical and experimental techniques, the derivation of realistic low-energy interactions between $NJ/\psi$ and $N\eta_{c}$ has been achieved by the lattice HAL QCD Collaboration~\cite{Ishii2007,ishii2012,aoki2013,sasaki2020,etminan2024prd}, based on $\left(2+1\right)$-flavor configurations with a pion mass close to the physical value, $m_{\pi} = 146$ MeV~\cite{yan2022prd}. It has been found that the potentials for $NJ/\psi$ and $N\eta_{c}$ are attractive at all distances and exhibit a characteristic long-range tail, which is consistent with the two-pion exchange potential.

Therefore, motivated by the above discussions, the $ \alpha\textrm{-}c\bar{c} $ correlation function is explored in this work, with the aim of probing the nature of $ N\textrm{-}c\bar{c} $ interactions as an independent source of information. The objective of this study is to provide an illustration of the potential insights that can be gained from the measurement of $ \alpha\textrm{-}c\bar{c} $ correlations. Since this is an exploratory investigation, simple techniques are employed. Given that the $ \alpha $-cluster is strongly bound and unlikely to undergo property changes, an effective $ \alpha \textrm{-} c\bar{c} $ nuclear potential is estimated through the single-folding method of the nucleon density within the $\alpha$-particle and the $ N\textrm{-}c\bar{c} $ interaction~\cite{Satchler1979,10.1143/ptp.117.251,Etminan:2019gds}.

The potentials of alpha-charmonium are evaluated using various well-established matter distributions and the root-mean-square (rms) radius of $\textrm{\ensuremath{^{4}}He}$ such as  central depression (CD) and the simple single Gaussian (SG) density. 
Next, the $\alpha\textrm{-}c\bar{c}$ potential obtained is fitted to an analytical Woods-Saxon (WS) type function. Subsequently, the Schrödinger equation is solved using the specified $\alpha\textrm{-}c\bar{c}$ potential as input, in order to calculate the binding energy, scattering parameters, and the $\alpha\textrm{-}c\bar{c}$ momentum correlation functions.

The organization of the paper is as follows:
In Sec.~\ref{sec:folding-Model}, the HAL QCD $N\textrm{-}c\bar{c}$ potentials are introduced and a brief description of the SFP approach is provided.
In Sec.~\ref{sec:Two-particle-CF}, the formalism for two-particle momentum correlation functions is succinctly reviewed. 
The results and discussions concerning $ \alpha\textrm{-}c\bar{c} $ are presented in Sec.~\ref{sec:result}. 
Finally, the summary and conclusions are given in Sec.~\ref{sec:Summary-and-conclusions}. 
\section{$ N\textrm{-}c\bar{c} $ interactions and $ \alpha\textrm{-}c\bar{c} $ SFP } \label{sec:folding-Model}
\subsection{State-of-the-art QCD nucleon-charmonium interactions}
The HAL QCD Collaboration presented a realistic lattice QCD simulations on the $ S $-wave $ N\textrm{-}c\bar{c} $ potentials, i.e.,  
$ NJ/\psi\left(^{4}S_{3/2}\right) $, $ NJ/\psi\left(^{2}S_{1/2}\right) $,
and $ N\eta_{c}\left(^{2}S_{1/2}\right) $~\cite{yan2022prd} close to the physical point $ m_{\pi} =146.4\left(4\right)$ MeV
(the notation $^{2s+1}L_{J}$ is used, $s$ is the total spin, $L$ and $J$ are orbital and total angular momentum).  
The $\left(2+1\right)$-flavor gauge configurations are generated on a large lattice volume of 
$\simeq\left(8.1\:\textrm{fm}\right)^{3}$ at the imaginary-time slices $ t/a=14 $ where $ a=0.0846 $ fm is the lattice spacing.

\begin{figure*}[hbt!]
	\centering
	\includegraphics[scale=1.0]{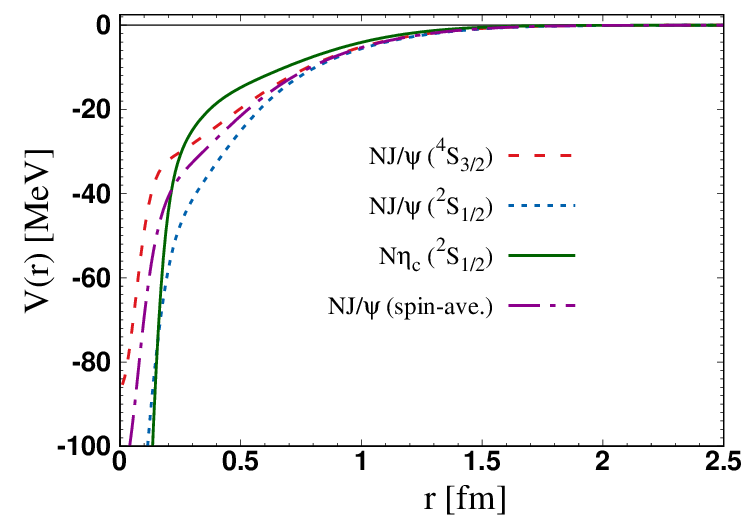}
	\caption{
		The $ S $-wave $ N\textrm{-}c\bar{c} $ potentials (in Eq.~\eqref{eq:NCharm_pot}) as functions of the distance between $ N $ and $ c\bar{c} $ are shown at the imaginary-time distances $ t/a=14$ by parametrization from Ref.~\cite{LyuPLB2025} as given in Table~\ref{tab:Charm-N-para}. 
		The spin-$3/2$ $ NJ/\psi $ is depicted by dashed red line, spin-$1/2$ $ NJ/\psi$ by dotted blue line, spin-$ 1/2 $ $N\eta_{c}$ by solid green line and the spin-averaged $ NJ/\psi $ (Eq.~\eqref{eq:NJpsi-spin-ave}) by dash-dotted purple line. 
		\label{fig:CharmN} }
\end{figure*}

In Ref.~\cite{LyuPLB2025}, for the imaginary-time slices $ t/a=14 $ an uncorrelated fit is performed with phenomenological three range Gaussians,
\begin{equation}
	V_{N \textrm{-}c\bar{c}}\left(r\right)=-\sum_{i=1}^{3}\alpha_{i}\exp\left[-\left(\frac{r}{\beta_{i}}\right)^{2}\right], \label{eq:NCharm_pot}
\end{equation} 
in the range of $ 0\le r\le 1.8 $ fm. The Gaussian functions describe the short-range attractive behavior.
The discrete lattice results are fitted $\chi^{2}/d.o.f\simeq 0.4, 0.4 $ and $ 0.6 $ for $ NJ/\psi\left(^{4}S_{3/2}\right) $, $ NJ/\psi\left(^{2}S_{1/2}\right) $,
and $ N\eta_{c}\left(^{2}S_{1/2}\right) $, respectively. 
The fitting parameters are taken directly from Ref.~\cite{LyuPLB2025} and given in Table~\ref{tab:Charm-N-para}.
The results of fit are shown in Fig.~\ref{fig:CharmN}.
It is found that the $ N\textrm{-}c\bar{c} $ potential is attractive at all distances and
have a characteristic long-range tail according to the two-pion exchange potential.
Nevertheless, the HAL QCD $ N\textrm{-}c\bar{c} $ interactions do not support the $ N\textrm{-}c\bar{c} $ bound state~\cite{PhysRevLett.127.172301,yan2022prd,Liang-ZhenPRD2025}.
Unlike the existence of a repulsive core in the nucleon interactions, 
the $ N\textrm{-}c\bar{c} $ interaction is expected to lack a repulsive core due to 
the Pauli exclusion principle does not act between common quarks~\cite{etminan2014,yan2022prd}.
The corresponding $ S $-wave scattering lengths are $ 0.30(2) $ fm, 
$ 0.38(4) $  fm, and $ 0.21(2) $ fm for $ NJ/\psi\left(^{4}S_{3/2}\right) $, $ NJ/\psi\left(^{2}S_{1/2}\right) $,
and $ N\eta_{c}\left(^{2}S_{1/2}\right) $, respectively. 
The HAL QCD results are larger than those from the photoproduction experiments assuming the vector meson dominance~\cite{PhysRevD.108.054018}. 

Since, the measurement of the two-body correlation functions and scattering parameters are currently limited to spin averaged quantities, 
the spin-averaged $ NJ/\psi $ potential is defined as
\begin{equation}
	NJ/\psi^{\textrm{spin-ave}}=\frac{2\:\:NJ/\psi\left(^{4}S_{3/2}\right)+NJ/\psi\left(^{2}S_{1/2}\right)}{3}. \label{eq:NJpsi-spin-ave}
\end{equation}
Figure~\ref{fig:CharmN} discloses a qualitative distinction among the  low energy spin-dependent $ N\textrm{-}c\bar{c} $ potentials, 
thus, it is always useful to study how these differences manifest themselves in the $\alpha\textrm{-}c\bar{c} $ two-particle momentum correlation functions.

\begin{table}[hbt!]
	\centering
	\caption{
		Fitting parameters in Eq.~\eqref{eq:NCharm_pot} with statistical errors in the parentheses at $ t/a=14 $.
		$ \alpha_{i} $ (in MeV) and $\beta_{i}$ (in fm) are taken directly from Ref.~\cite{LyuPLB2025}.
		\label{tab:Charm-N-para}}	
	\begin{tabular}{ccccccc}
		\hline
		\hline 
		& $\alpha_{1}$ & $\beta_{1}$& $\alpha_{2}$&$\beta_{2}$&$\alpha_{3}$&$\beta_{3}$\\
		\hline
		$ NJ/\psi\left(^{4}S_{3/2}\right) $& $ 51(1)$  & $ 0.09(1) $ & $ 13(6)$ & $ 0.49(7)$ & $22(5)$& $0.82(6)$ \\	
		$ NJ/\psi\left(^{2}S_{1/2}\right) $& $ 101(1) $& $ 0.13(1) $ & $33(6)$  & $ 0.44(5)$ & $23(8)$& $0.83(9)$ \\
		$N\eta_{c}\left(^{2}S_{1/2}\right)$& $ 264(1) $& $ 0.11(1) $ & $28(13)$ & $0.24(6)$  & $22(2)$& $0.77(3)$ \\
		\hline
		\hline 	
	\end{tabular}
\end{table}
\subsection{$ \alpha\textrm{-}c\bar{c} $ SFP} \label{subsec:sfp}
The effective $ \alpha \textrm{-} c\bar{c} $ nuclear potential is approximated by the SFP model
\begin{equation}
	U_{\alpha\textrm{-}c\bar{c}}\left(r\right)=\int\rho\left(x\right)V_{N\textrm{-}c\bar{c}}\left(\left|\textbf{r}-\textbf{x}\right|\right)d\textbf{x},\label{eq:V_alfaOmega}
\end{equation}		
where $V_{N\textrm{-}c\bar{c}}\left(\left|\textbf{r}-\textbf{x}\right|\right)$ is $N\textrm{-}c\bar{c}$ potential between
the nucleon at $\textbf{x}$ and the charmonium at $\textbf{r}$~\cite{Satchler1979, Etminan:2019gds}; 
moreover, $\rho\left(x\right)$ is the nucleon density function in
$\alpha$-particle as (NNNN) cluster with $J^{\pi}=0^{+}$ at a distance $\textbf{x}$ from its
center-of-mass.

When the distances between \(\alpha\) and \(c\bar{c}\) are sufficiently large, the clustering can be described as \(c\bar{c} + (NNNN)\). However, inside and near the \(\alpha\) cluster, all possible five-body configurations of the clustering should be considered. 
Since no bound states are formed by the \(c\bar{c}\) meson in any subsystem, it is assumed that the \(c\bar{c} + (NNNN)\) clusterization is dominant, and therefore the folding potential is regarded as an applicable approach for the \(c\bar{c}\textrm{-}\alpha\) interaction.

 The information that the (NNNN) cluster possesses spin-0 and positive parity is implemented by employing the cluster’s intrinsic \(0^{+}\) density, which corresponds to the alpha-like state.
 In practice, the potential \(U_{\alpha\textrm{-}c\bar{c}}(r)\) is obtained by integrating the effective interaction \(V_{N\textrm{-}c\bar{c}}\) over the internal coordinates of the cluster, weighted by its internal density or by its overlap with the relevant internal states.
 If (NNNN) is constrained to have total spin \(S=0\) and positive parity \(\pi=+\), the internal wave function of the four-nucleon cluster is fixed to a specific spatially symmetric configuration. In an alpha-like cluster, the intrinsic wave function is highly symmetric and antisymmetric with respect to fermionic exchange, yielding \(J^{\pi}=0^{+}\).

 Furthermore, respect must be accorded to the Pauli principle by the folding formalism between the nucleons of the cluster and the external hadrons.
  Unlike the nucleon–nucleon interaction, where a repulsive core is present, the absence of a repulsive core in the \(N\textrm{-}c\bar{c}\) interaction is partially attributed to the fact that the nucleon \(N\) and the charmonium \(c\bar{c}\) do not share valence quarks, so the Pauli exclusion principle between common quarks does not operate. Consequently, Pauli-blocking corrections are not required here.
  
  The folding procedure has been successfully applied in Refs.~\cite{Filikhin_2005, HiyamaPRC2022, filikhin2024phihe, etminanPRCphiNalpha2025} for analyses of the \(\Lambda NN\alpha,\) \(\Xi N\alpha\alpha,\) \(\phi\alpha\alpha,\) \(\phi N\alpha\) systems, and in Ref.~\cite{filikhin2000alpha8be}, the first \(0_{2}^{+}\) excited state of \({}^{12}\)C is described on the basis of the Faddeev equation. Agreement with experimental data and with calculations conducted within various theoretical frameworks has been testified by the predictions of these works.
\subsubsection{Approximation for when charmonium is outside of alpha}
As the first and very simple case, it is supposed that the charmonium is just near and outside of alpha particle. 
Therefore, the integral in Eq.~\eqref{eq:V_alfaOmega} is solved by the assumption that $\left|\textbf{r}-\textbf{x}\right|\geqslant1.9$ fm.
This is a conservative approximation and emphasizes that at large distances between $c\bar{c}$ and $\alpha$, the clustering is described as $c\bar{c}+\left(NNNN\right)$,  and gives the minimum of interaction.

For this case, the nucleon density function in $\alpha$-particle is taken to be~\cite{Akaishi1986},
\begin{equation}
	\rho\left(r\right)=4\left(\frac{4\beta}{3\pi}\right)^{3/2}\exp\left(-\frac{4}{3}\beta r^{2}\right),\label{eq:nucleon-density}
\end{equation}
$\beta$ is a constant and it is defined by the rms radius of $\textrm{\ensuremath{^{4}}He}$, i.e,
$ rms={3}/{\sqrt{8\beta}}=1.47 $ fm~\cite{Akaishi1986}. 

Thanks to the three-Gaussian analytical fitting of the HAL QCD potential, the folding integral can be solved analytically for the Gaussian matter distribution of the \(\alpha\) cluster (see the Appendix of Refs.~\cite{filikhin2024phihe,FilikhinSPC2025}).

\subsubsection{Central depression density distribution}
In Ref.~\cite{SaitoPRC1997} suggested parameterization of $^{4}$He density in the form 
\begin{equation}
	{\rho}_{CD}\left(r\right)= \rho_{0} \left(1+\gamma r^{2}\right)e^{-\lambda r^{2}}, \label{eq:ctr_depRHO}
\end{equation}
that reproduces the central depression measured in the charge density
and gives $rms=\sqrt{\frac{3\left(5\gamma+2\lambda\right)}{2\lambda\left(3\gamma+2\lambda\right)}}=1.56$
fm. The value of the parameters is, ${\rho}_{0} = 0.04775$ fm$^{-3}$, $\gamma=1.34215$
fm$^{-2}$, and $\lambda=0.904919$ fm$^{-2}$. Substitution of the HAL QCD $V_{N\textrm{-}c\bar{c}}$ interaction
parametrized with three Gaussian functions Eq.~\eqref{eq:NCharm_pot} and density distribution
function in Eq.~\eqref{eq:ctr_depRHO} gives~\cite{filikhin2024phihe}
\begin{equation}
	V_{\alpha\textrm{-}c\bar{c}}\left(r\right)=2\pi^{3/2} {\rho}_{0} \sum_{i=1}^{3}\frac{\alpha_{i}\beta_{i}^{3}}{c_{i}^{7/2}}\exp\left(-\frac{\lambda}{c_{i}}r^{2}\right)\left[\lambda\left(3\gamma+2\lambda\right)\beta_{i}^{4}+\left(3\gamma+4\lambda\right)\beta_{i}^{2}+2\gamma r^{2}+2\right], \label{eq:anal_VccbarAlpha}
\end{equation}
where $c_{i}=1+\lambda\beta_{i}^{2}$.

\subsubsection{Simple single Gaussian density distribution} 
The single folding potential is found to be sensitive to the rms radius of \({}^{4}\)He. Recently, the rms matter radius of \({}^{4}\)He has been measured to be \(1.70 \pm 0.14\) fm~\cite{PhysRevC.109.L012201}.
From these analyses, the rms charge radius of \({}^{4}\)He is smaller than the rms matter radius. However, the values of the rms charge and matter
radii are within the statistical errors. 
The influence of the rms radius on the obtained \(\phi \alpha\) potential has been investigated in Refs.~\cite{filikhin2024phihe, ETMINAN2025PLB139564}.
While this is a significant puzzle, in present calculation of the folding potential, the densities that reproduce the rms radii \(1.70 \pm 0.14\) fm~\cite{PhysRevC.109.L012201}, 1.56, 1.70, and 1.84 fm are used.

As follows from Ref.~\cite{PhysRevC.109.L012201}, the simple single Gaussian (SG) matter distribution model 
\begin{equation}
	{\rho}_{G}\left(r\right)=\rho_{0} \exp\left(-\lambda r^{2}\right),\label{eq:gauss-dist}
\end{equation}
with $\rho_{0} = \left(\frac{\lambda}{\pi}\right)^{3/2}$ gives $\left\langle r^{2}\right\rangle ^{1/2} = \sqrt{\frac{3}{2\lambda}} =1.56, 1.70, 1.84 $ fm and describes the experimental data with parameters from Ref.~\cite{PhysRevC.109.L012201}. 
Substitution of the HAL QCD $V_{N\textrm{-}c\bar{c}}$ interaction parametrized with three Gaussian functions as in Eq.~\eqref{eq:NCharm_pot} and the density distribution function in Eq.~\eqref{eq:gauss-dist} is also found to yield Eq.~\eqref{eq:anal_VccbarAlpha}, where the parameter $\gamma=0.0$ in this case.

\section{Two-particle correlation function formalism}\label{sec:Two-particle-CF}
The formalism of the two-particle correlation function has been extensively discussed in numerous studies~\cite{KOONIN197743,Pratt1986,PhysRevC.91.024916,OHNISHI2016294,cho2017exotic,bazak2020production}, with only the essential formulas being provided here.  

The general form of the $\alpha\textrm{-}c\bar{c}$ correlation function is determined by the formation mechanism of the particles, which depends on whether $\alpha$ nuclei are emitted from a source alongside other hadrons or are formed through final-state interactions among emitted nucleons.  
In the first scenario, the $\alpha$ nucleus can be treated as a point-like particle, while in the second scenario, it is regarded as a bound state of four nucleons~\cite{bazak2020production}. If the $\alpha$ nucleus is modeled as a point-like particle emitted from a source, the $\alpha\textrm{-}c\bar{c}$ correlation function is expressed as 
\begin{equation}
	\frac{dP_{\alpha\textrm{-}c\bar{c}}}{d^{3}p_{\alpha}d^{3}p_{c\bar{c}}}=C\left(\boldsymbol{p}_{\alpha},\boldsymbol{p}_{c\bar{c}}\right)\frac{dP_{\alpha}}{d^{3}p_{\alpha}}\frac{dP_{c\bar{c}}}{d^{3}p_{c\bar{c}}},
\end{equation}
where $\frac{dP_{\alpha}}{d^{3}p_{\alpha}}$, $\frac{dP_{c\bar{c}}}{d^{3}p_{c\bar{c}}}$, and $\frac{dP_{\alpha\textrm{-}c\bar{c}}}{d^{3}p_{\alpha}d^{3}p_{c\bar{c}}}$ represent the probability densities for measuring $\alpha$, $c\bar{c}$, and the $\alpha\textrm{-}c\bar{c}$ pair with momenta $\boldsymbol{p}_{\alpha}$, $\boldsymbol{p}_{c\bar{c}}$, and $\left(\boldsymbol{p}_{\alpha},\boldsymbol{p}_{c\bar{c}}\right)$, respectively.  

If the correlation arises from final-state interactions, the correlation function is defined using the Koonin-Pratt (KP) formula~\cite{Pratt1986,PhysRevC.91.024916,OHNISHI2016294},
\begin{equation}
	C\left(q\right)=1+\int_{0}^{\infty}4\pi r^{2}\:dr\:S_{r}\left(r\right)\left[\left|\psi\left(q,r\right)\right|^{2}-\left|j_{0}\left(qr\right)\right|^{2}\right], \label{eq:kp}
\end{equation}      
where $S_{r}\left(r\right)=\exp\left(-\frac{r^{2}}{4R^{2}}\right)/\left(4\pi R^{2}\right)^{3/2}$ is the single-particle source function,
assumed to be a spherical, static Gaussian with radius $R$ and describes the distribution of
$\alpha\textrm{-}c\bar{c}$ pair production at the relative distance $r$. 
When considering source sizes $R_{c\bar{c}}$ and $R_{\alpha}$ for the $c\bar{c}$ and $\alpha$ emissions, respectively,
an effective source radius $R$ is calculated as $R=\sqrt{\left(R_{c\bar{c}}^{2}+R_{\alpha}^{2}\right)/2}$.
The $j_{l=0}\left(qr\right)=\sin\left(qr\right)/qr$ is the spherical Bessel function  
and $\psi\left(q,r\right)$ is the S-wave scattering wave function, which is solutions to the Schr\"{o}dinger equation for the given two-body $\alpha\textrm{-}c\bar{c}$ potential.

The computation of hadron-light nuclei correlation functions raises the question of how the source function for light nuclei should be selected. As investigated in Refs.~\cite{mrowczynski2019hadron, MROWCZYNSKI2025139413}, the proton-deuteron correlation function exhibits distinct forms depending on whether deuterons are emitted directly from the fireball alongside other hadrons or are formed later through final-state interactions.  

In these two scenarios, the source functions of deuterons, which are incorporated into the formulas for the $p$-deuteron correlation function, differ. As a result, the source radii derived from the correlation functions vary by a factor of $\sqrt{4/3}$. A similar situation arises when the $p\textrm{-}^3$He correlation function is analyzed, though the computation becomes more complex due to the involvement of a four-body problem.  

It has been demonstrated in Ref.~\cite{bayegan2008three} that if $^3$He is assumed to be emitted directly from the fireball, the source radius inferred from the correlation function is reduced by a factor of $\sqrt{3/2}$ compared to the scenario where nucleons emitted from the fireball coalesce into $^3$He via final-state interactions.  
The selection of source sizes for $\alpha\textrm{-}c\bar{c}$ correlations is addressed in Section~\ref{sec:result}.  

In the case where the source size significantly exceeds the interaction range,
the wave function's asymptotic form, $\psi\left(q,r\right)\rightarrow j_{0}\left(qr\right)+f\left(q\right)\exp\left(iqr\right)/r$,
simplifies the correlation function evaluation, resulting in the Lednicky-Lyuboshits (LL) approximation~\cite{Lednicky:1981su},
\begin{equation}
	C_{LL}\left(q\right)=1+\frac{\left|f\left(q\right)\right|^{2}}{2R^{2}}F_{0}\left(\frac{r_{0}}{R}\right)+\frac{2\textrm{Re}\:f\left(q\right)}{\sqrt{\pi}R}F_{1}\left(2qR\right)-\frac{\textrm{Im}\:f\left(q\right)}{R}F_{2}\left(2qR\right), \label{eq:ll}
\end{equation}   
The scattering amplitude, which is approximated as  
$f(q) \approx \frac{1}{\left(-1/a_0 + r_0 q^2/2 - i q\right)}, $  
where $a_0$ represents the scattering length and $r_0$ denotes the effective range, is derived using the effective range expansion (ERE) formula, as presented in Eq.~\eqref{eq:ERE}.  
Additionally, the functions  
$ F_1(x) = \int_{0}^{x} dt \, \frac{e^{t^2 - x^2}}{x}, \quad F_2(x) = \frac{1 - e^{-x^2}}{x}, $  
are defined, and the correction term  
$ F_0(x) = 1 - \frac{x}{2\sqrt{\pi}} $  
is introduced to account for deviations of the asymptotic form from the true wave function, as discussed in Refs.~\cite{Lednicky:1981su,OHNISHI2016294}.  

\section{Numerical results {and discussion} } \label{sec:result}
\begin{figure*}[hbt!]
	\centering
	\includegraphics[scale=0.62]{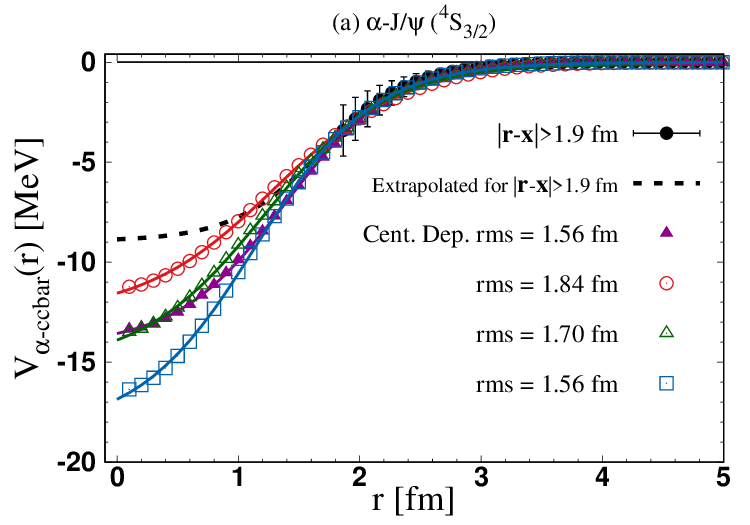}\includegraphics[scale=0.62]{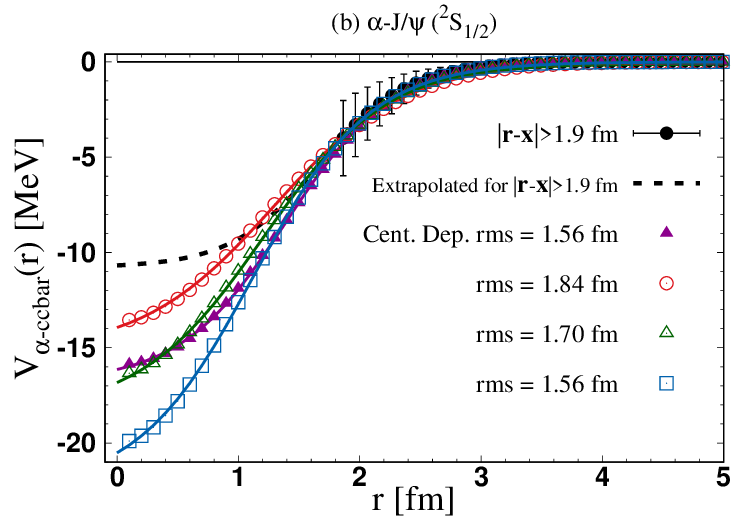} ~
	\includegraphics[scale=0.62]{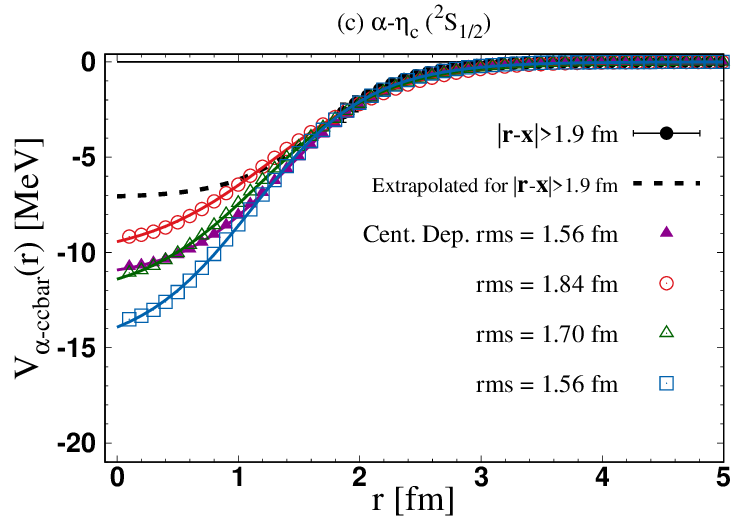} \includegraphics[scale=0.62]{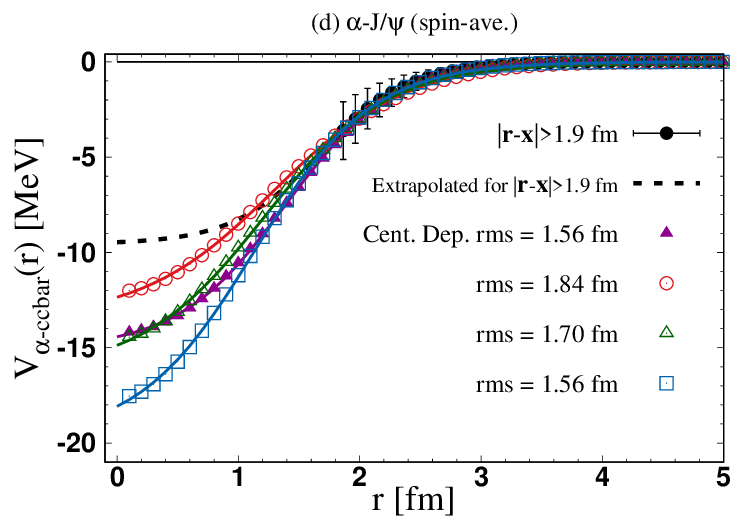}
	\caption{ 
		The data points show the obtained $ \alpha\textrm{-}c\bar{c} $ potentials through solving integral equation and the solid line display the corresponding WS fits for
		(a) spin-$3/2$ $ NJ/\psi $, (b) spin-$1/2$ $ NJ/\psi$, (c) spin-$ 1/2 $ $N\eta_{c}$ and 
		(d) the spin-averaged $ NJ/\psi $.  
		The different symbols correspond to the different calculation method and rms radii.
		The data points (black full   circles) show the $U_{\alpha\textrm{-}c\bar{c}}\left(r\right)$  that calculated by Eq.~\eqref{eq:V_alfaOmega} 
		using  density function which is given by Eq.~\eqref{eq:nucleon-density}. 
		The corresponding errors for data points are statistical.
		The obtained potential using Eq.~\eqref{eq:anal_VccbarAlpha}  and central depression  density distribution in Eq.~\eqref{eq:ctr_depRHO} 
		which gives  rms matter radius of $^4$He, 1.56 fm is shown by the full  purple triangle.
		The analytical obtained potential via Eq.~\eqref{eq:anal_VccbarAlpha}  using simple single Gaussian density distribution (Eq.~\eqref{eq:gauss-dist})  
		for the values of rms radius: 1.56, 1.70, and 1.84 fm, are indicated by the hollow red circle, green triangle and blue square, respectively.
		The  lines show the fitting of $ \alpha\textrm{-}c\bar{c}$.
		The results of the fit are presented in Table~\ref{tab:Charm-alpha-para}. 
		\label{fig:v_CharmAlpha}
	}
\end{figure*}

The  behavior of the obtained single-folding potentials, $ \alpha\textrm{-}J/\psi\left(^{4}S_{3/2}\right) $, $ \alpha\textrm{-}J/\psi\left(^{2}S_{1/2}\right) $, 
$ \alpha\textrm{-}\eta_{c}\left(^{2}S_{1/2}\right) $ and the spin-averaged $ \alpha\textrm{-}J/\psi $ (spin-ave.) are shown in Fig.~\ref{fig:v_CharmAlpha} (a), (b), (c) and (d), respectively. 
These potentials are obtained through solving the Eq.~\eqref{eq:V_alfaOmega} for $ NJ/\psi\left(^{4}S_{3/2}\right) $, $ NJ/\psi\left(^{2}S_{1/2}\right) $, $ N\eta_{c}\left(^{2}S_{1/2}\right) $ and the spin-averaged $ NJ/\psi $ interactions.
For each interaction, the $ \alpha\textrm{-}c\bar{c} $ potential has been calculated with different density distribution functions with different rms radii as described in subsection~\ref{subsec:sfp}.

In each panel of Fig.~\ref{fig:v_CharmAlpha}, the data points show the $U_{\alpha\textrm{-}c\bar{c}}\left(r\right)$  that calculated by  Eq.~\eqref{eq:V_alfaOmega} using  density function which is given by Eq.~\eqref{eq:nucleon-density}. 
In this case, the integral in Eq.~\eqref{eq:V_alfaOmega} is solved by the assumption that $\left|\textbf{r}-\textbf{x}\right|\geqslant1.9$ fm.
This is a conservative approximation and emphasizes that at large distances between $c\bar{c}$ and $\alpha$, the clustering is described as $c\bar{c}+\left(NNNN\right)$.

The obtained analytical potential as provided by Eq.~\eqref{eq:anal_VccbarAlpha}  employing the central depression  density distribution (Eq.~\eqref{eq:ctr_depRHO}) 
which gives rms matter radius of $^4$He, 1.56 fm is shown by the full  purple triangle in Fig.~\ref{fig:v_CharmAlpha}.
Also, the derived analytical potentials via Eq.~\eqref{eq:anal_VccbarAlpha}  using simple single Gaussian density distribution (Eq.~\eqref{eq:gauss-dist})  
for the values of rms radius: 1.56, 1.70, and 1.84 fm, are presented in this figure.
The  lines shows the Woods-Saxonian fitting of $ \alpha\textrm{-}c\bar{c}$  
by using analytical form function,$  U_{\alpha\textrm{-}c\bar{c}}^{fit}\left(r\right) $ given in Eq.~\eqref{eq:ws-fit}. 
Moreover, the results of the fit, corresponding binding energy and the scattering parameters are presented in Table~\ref{tab:Charm-alpha-para}.

One can see that the depth of the $	U_{\alpha\textrm{-}c\bar{c}}^{fit}$ potential is very sensitive
to the value of the rms radius, and the selected density distribution, approximately varying in the interval $\left[-9,-17\right]$, $\left[-11,-21\right]$, $\left[-7,-14\right]$ and $\left[-10,-18\right]$ MeV for $ \alpha\textrm{-}J/\psi\left(^{4}S_{3/2}\right) $, $ \alpha\textrm{-}J/\psi\left(^{2}S_{1/2}\right) $, 
$ \alpha\textrm{-}\eta_{c}\left(^{2}S_{1/2}\right) $ and $ \alpha\textrm{-}J/\psi $ (spin-ave.) potentials, respectively.

Also, it can be interesting to compare two different density functions of $ ^{4}\textrm{He}$, CD (Eq.~\eqref{eq:ctr_depRHO}) and simple SG (Eq.~\eqref{eq:gauss-dist}) density, with the same rms radius of 1.56 fm.  The present results as provided by Fig.~\ref{fig:v_CharmAlpha}  show that the $\alpha\textrm{-}c\bar{c}$ potentials  extracted from the simple SG density distribution function  is in almost all cases deeper than the corresponding one obtained from the CD density matter model.

\begin{figure*}[hbt!]
	\centering
	\centering
	\includegraphics[scale=0.62]{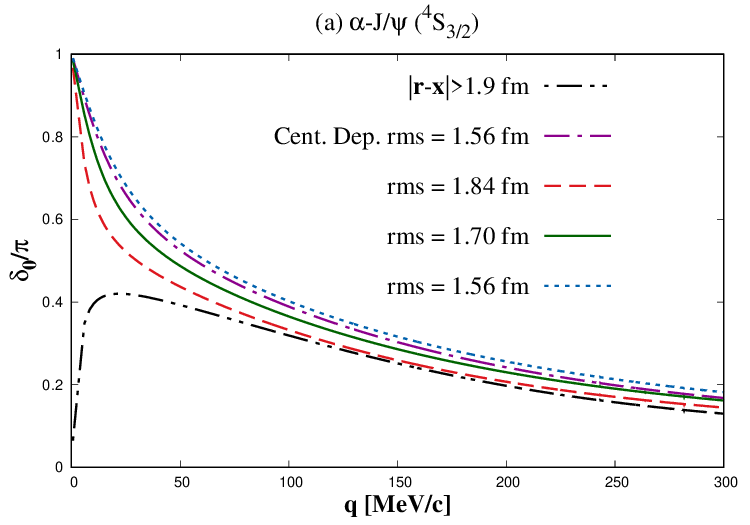}\includegraphics[scale=0.62]{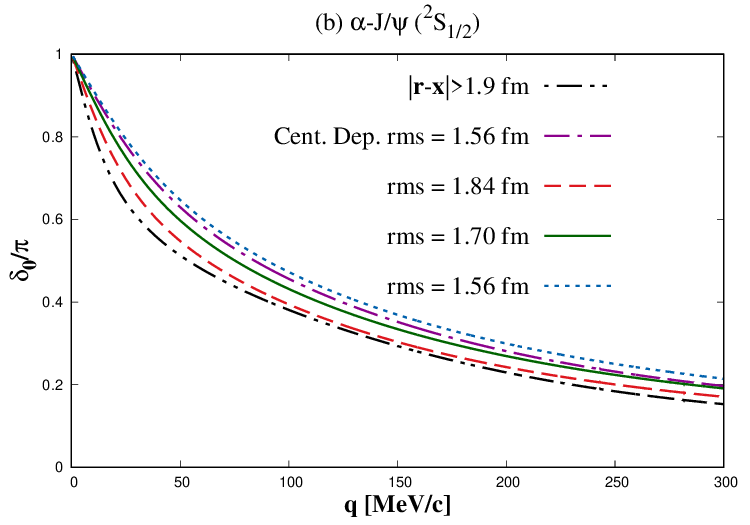} ~
	\includegraphics[scale=0.62]{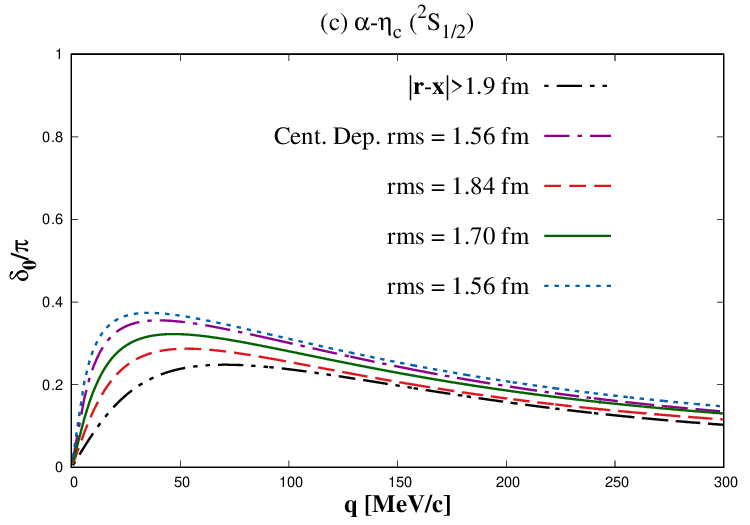} \includegraphics[scale=0.62]{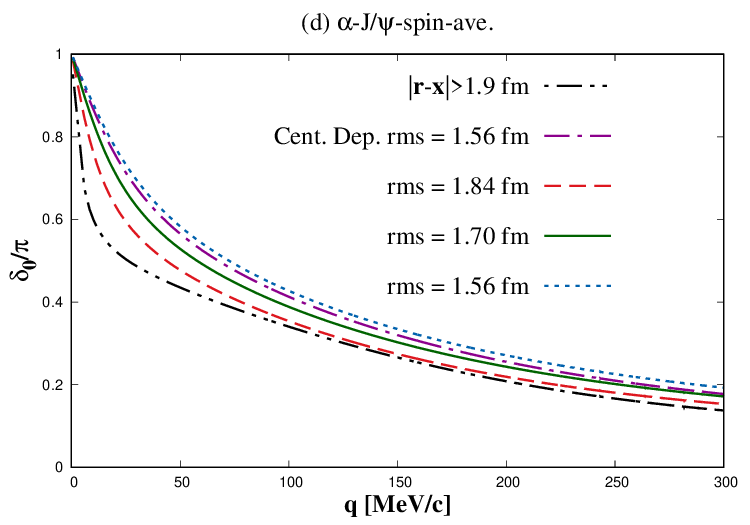}
	\caption{
		The phase shifts $\delta_{0} / \pi$ for the $\alpha\textrm{-}c\bar{c}$ system are depicted as functions of the relative momentum $q = \sqrt{2\mu E}$, based on the obtained potential $U_{\alpha\textrm{-}c\bar{c}}(r)$. Here, $\mu$ represents the reduced mass of the $\alpha\textrm{-}c\bar{c}$ system. 
		\label{fig:phase_DG_HAL}}
\end{figure*}

For practical applications and computing scattering phase shifts, correlation functions and binding energies,
$U_{\alpha\textrm{-}c\bar{c}}$ is fitted to a Woods-Saxon form (motivated by Dover-Gal model of potential~\cite{dover1983})
\begin{equation}
	U_{\alpha\textrm{-}c\bar{c}}^{fit}\left(r\right)=-U_{0}\left[1+\exp\left(\frac{r-R_{c}}{c}\right)\right]^{-1} , \label{eq:ws-fit}
\end{equation}
the parameters $ U_{0}$ is the depth, $ R_{c} = r_{c} A^{1/3} $ the radius of the nucleus (here $ A = 4 $) and $ c $ is the surface diffuseness. 
The fit parameters are listed in Table~\ref{tab:Charm-alpha-para} for the obtained $\alpha\textrm{-}c\bar{c}$ potentials based on different $N\textrm{-}c\bar{c}$ spin channels. And, the fitting potentials are depicted in Fig.~\ref{fig:v_CharmAlpha} by lines. 
Using these fitted potentials, the Schr\"{o}dinger equation is solved to extract binding energies and scattering observables; 
the phase shifts calculated from these potentials is shown in Fig.~\ref{fig:phase_DG_HAL} and  indicate an attractive interaction for all cases.

According to the data provided in Table~\ref{tab:Charm-alpha-para}, the highest binding energy, which is about 0.6 MeV, occurs for the spin-$1/2$ $\alpha\textrm{-}J/\psi$ system with an rms radius of 1.56 fm with the simple SG matter density model, followed by the same system with a value of 0.5 MeV with the CD matter density model. The spin-$3/2$ and spin averaged $\alpha\textrm{-}J/\psi$ systems could form a loosely bound state approximately in the range of 0.1-0.3  MeV for central binding energy,  while $ \alpha\textrm{-}\eta_{c}\left(^{2}S_{1/2}\right) $ system remain unbound relative to the $\alpha + c\bar{c}$ threshold. In general, as expected, the energy dependence of the corresponding system decreases with increasing rms radius.

In Ref.~\cite{COBOSMARTINEZ2020135882}, $ \eta_{c}\textrm{-}^{4}\textrm{He} $ bound state energies $ 1.49,3.11,5.49 $ and $ 8.55 $ MeV are calculated for different values of the cutoff parameter $ \Lambda_{D}=1500,2000,2500 $ and $ 3000 $ MeV, respectively. 
Where the main input, particularly the medium-modified $ D $ and $ D^{*} $ meson masses, 
besides the density distributions in nuclei are studied within the quark-meson coupling model.

The low-energy phase shift behavior in Fig.~\ref{fig:phase_DG_HAL} allows the extraction of scattering length and effective range via the effective range expansion (ERE) formula up to the next-leading-order,		
\begin{equation}
	q\cot\delta_{0}=-\frac{1}{a_{0}}+\frac{1}{2}r_{0}q^{2}+\mathcal{O}\left(q^{4}\right).\label{eq:ERE}
\end{equation}				
The calculated values for these quantities are presented in Table~\ref{tab:Charm-alpha-para}, for all models of interactions. 

\begin{table}[hbt!]
	\centering
	\caption{
		The fit parameters of $U_{\alpha\textrm{-}c\bar{c}}$ in Eq.~\eqref{eq:ws-fit}
		and the corresponding low-energy parameter, scattering length $ a_{0} $, effective range $r_{0}$ and binding energy $B_{\alpha\textrm{-}c\bar{c}}$, are given for  $ \alpha\textrm{-}c\bar{c} $ potentials.
		Calculations are done by using the experimental masses,
		i.e., $ m_{\alpha}=3727.38 $, $ m_{J/\psi}=3096.9 $ MeV/c and $ m_{\eta_{c}}=2984.1 $ MeV/c. 
		Furthermore, the results corresponding to $  c\bar{c} $ mass value derived by the lattice simulation~\cite{LyuPLB2025} are given within parentheses.  
		The experimental parameters for neutron-neutron scattering are given as \( (a_0, r_0) = (-18.5, 2.80) \) fm and are used for comparison.
		{Note that all displayed numbers represent central values.}
		\label{tab:Charm-alpha-para}}	
	\begin{tabular}{cccccccc}
		\hline
		\hline 
		$\alpha\textrm{-}c\bar{c}$ &rms radius (fm)& $ U_{0} $ (MeV)& $ r_{c} $(fm)& $ c $ (fm)& $a_{0}$(fm)&$r_{0}$(fm)& $B$(MeV) \\
		\hline
		\multirow{4}{*}{$ \alpha\textrm{-}J/\psi\left(^{4}S_{3/2}\right) $}&1.47& $ 8.95 $ & $ 1.061 $ & $ 0.367 $ & $ -70.0 $ & $ 2.30 $ & $-\left(-\right) $ \\	
		&1.56(CD)& $ 14.20 $ & $ 0.870 $ & $ 0.449 $ & $ 11.6 $ & $ 2.05  $ & $ 0.1$ \\	
		&1.56(SG)& $ 18.44 $ & $ 0.720 $ & $ 0.482 $ & $ 10.2 $ & $ 1.98  $ & $ 0.1$ \\
		&1.70    & $ 15.20 $ & $ 0.768 $ & $ 0.514 $ & $ 16.4 $ & $ 2.21  $ & $ 0.0$ \\
		&1.84    & $ 12.63 $ & $ 0.817 $ & $ 0.546 $ & $ 35.2 $ & $ 2.45  $ & $ -$ \\
		&&&&&&&\\						
		\multirow{4}{*}{$ \alpha\textrm{-}J/\psi\left(^{2}S_{1/2}\right) $}&1.47& $ 10.79 $& $ 1.051 $ & $ 0.366 $ & $ 13.0  $ & $ 2.1  $  & $0.1\left(0.1\right)$ \\
		&1.56(CD)& $ 16.80 $& $ 0.877 $ & $ 0.437 $ & $ 6.0 $ & $ 1.81 $  & $ 0.5 $ \\
		&1.56(SG)& $ 22.47 $& $ 0.710 $ & $ 0.479 $ & $ 5.5 $ & $ 1.74 $  & $ 0.6 $ \\
		&1.70    & $ 18.42 $& $ 0.760 $ & $ 0.511 $ & $ 7.0 $ & $ 1.95 $  & $ 0.3 $ \\
		&1.84    & $ 15.24 $& $ 0.810 $ & $ 0.543 $ & $ 9.3 $ & $ 2.16 $  & $ 0.2 $ \\
		&&&&&&&\\	
		\multirow{4}{*}{$ \alpha\textrm{-}\eta_{c}\left(^{2}S_{1/2}\right) $}&1.47&$ 7.12 $ & $ 1.056 $ & $ 0.360 $ & $ -5.4 $  & $ 2.7 $  & $-$ \\
		&1.56(CD)& $ 11.37 $& $ 0.875 $ & $ 0.434 $ & $ -20.1 $ & $ 2.37  $  & $ - $ \\
		&1.56(SG)& $ 15.23 $& $ 0.707 $ & $ 0.475 $ & $ -26.5 $ & $ 2.32  $  & $ - $ \\
		&1.70    & $ 12.47 $& $ 0.757 $ & $ 0.507 $ & $ -13.4 $ & $ 2.60 $  & $ -$ \\
		&1.84    & $ 10.31 $& $ 0.807 $ & $ 0.540 $ & $ -9.3  $ & $ 2.90  $  & $ -$ \\
		&&&&&&&\\
		\multirow{4}{*}{$ \alpha\textrm{-}J/\psi $ (spin-ave.)}&1.47& $ 9.56 $ & $ 1.057 $ & $ 0.367 $ & $ 52.2 $  & $ 2.2 $  & $-\left(-\right)$ \\ 
		&1.56(CD)& $ 15.06 $& $ 0.872 $ & $ 0.445 $ & $ 8.6   $ & $ 1.96 $  & $ 0.2$ \\
		&1.56(SG)& $ 19.78 $& $ 0.715 $ & $ 0.481 $ & $ 7.8   $ & $ 1.90  $  & $ 0.3$ \\
		&1.70    & $ 16.27 $& $ 0.765 $ & $ 0.513 $ & $ 11.0  $ & $ 2.11 $  & $ 0.1$ \\
		&1.84    & $ 13.50 $& $ 0.815 $ & $ 0.545  $ & $ 17.4 $ & $2.34 $  & $0.04$ \\
		\hline
		\hline 	%
	\end{tabular}
\end{table}

\begin{figure*}[hbt!]
	\centering
	\includegraphics[scale=0.64]{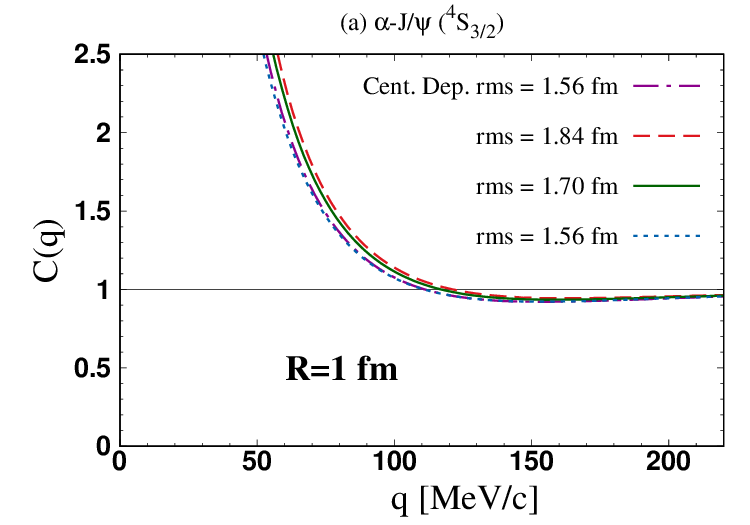}    
	
	\includegraphics[scale=0.64]{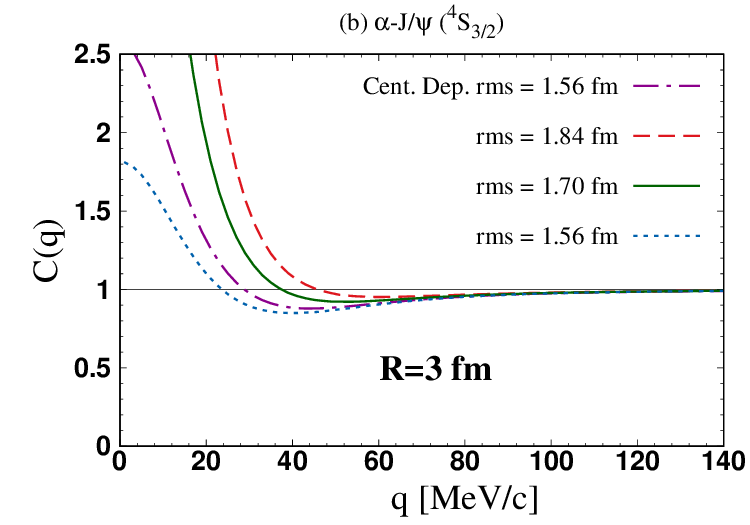} \includegraphics[scale=0.64]{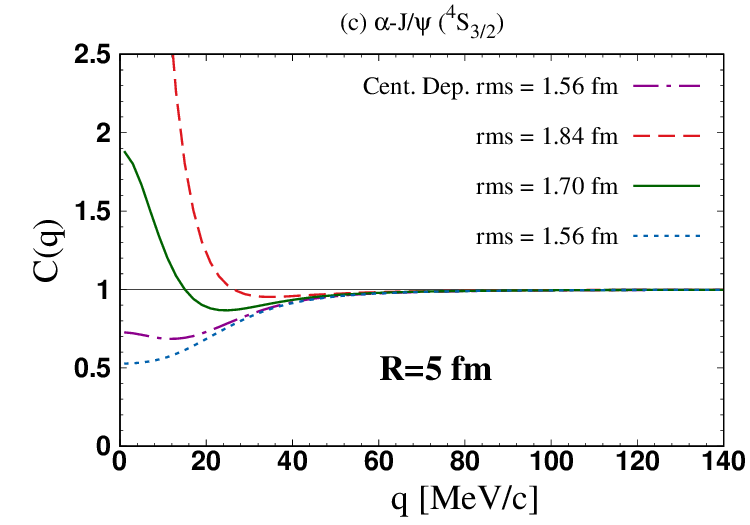}
	\caption{ The spin-$3/2$ $ \alpha\textrm{-}J/\psi $  correlation functions for three different source sizes: (a) $ R=1 $ fm, (b) $ R=3 $ fm and (c) $ R=5 $ fm, with models of central depression (Cent. Dep.) density distribution as given by  Eq.~\eqref{eq:ctr_depRHO} 
		which gives  rms matter radius of $^4$He, 1.56 fm (dash-dotted magenta line) and with the model of simple single Gaussian density distribution as given by Eq.~\eqref{eq:gauss-dist} for the three values of the rms radius: $ 1.84 $ fm (long dashed red line), $ 1.70 $ fm (solid green line), and $ 1.56 $ fm (dotted blue magenta line).
		\label{fig:cq-kp-JPsi4S32Alpha}   }
\end{figure*}

\begin{figure*}[hbt!]
	\centering
	\includegraphics[scale=0.64]{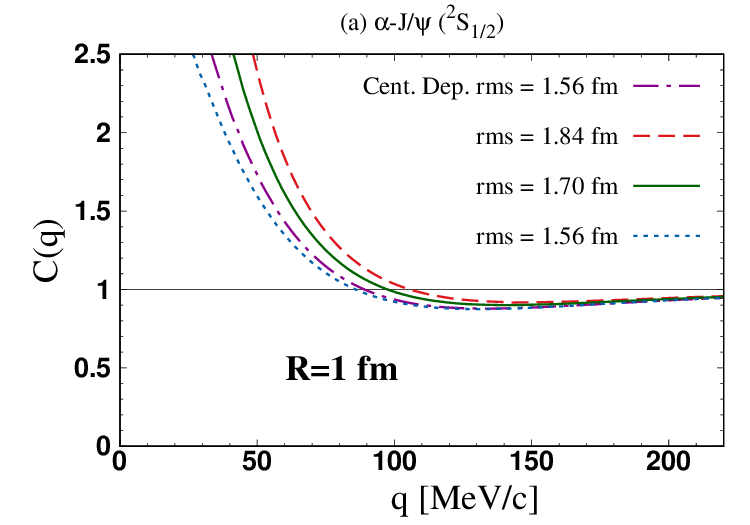}    
	
	\includegraphics[scale=0.64]{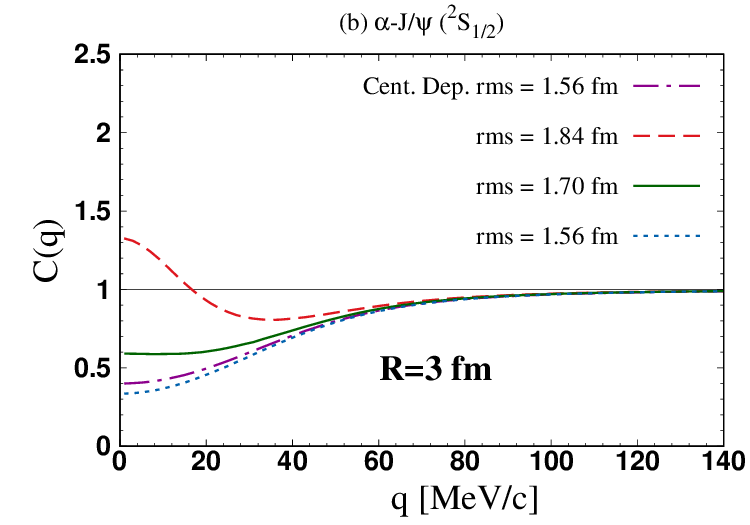} \includegraphics[scale=0.64]{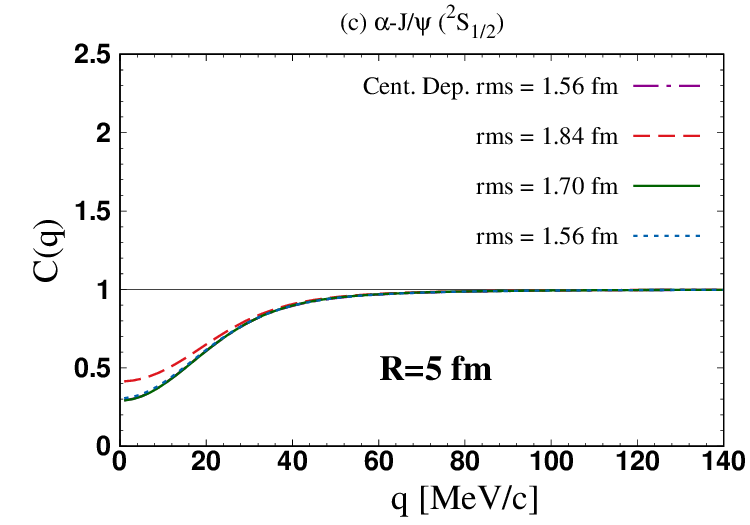}
	\caption{ The spin-$1/2$ $ \alpha\textrm{-}J/\psi $  correlation functions for three different source sizes: (a) $ R=1 $ fm, (b) $ R=3 $ fm and (c) $ R=5 $ fm. Symbols have the same description as in Fig.~\ref{fig:cq-kp-JPsi4S32Alpha}.
		\label{fig:cq-kp-JPsi2S12Alpha}   }
\end{figure*}

\begin{figure*}[hbt!]
	\centering
	\includegraphics[scale=0.64]{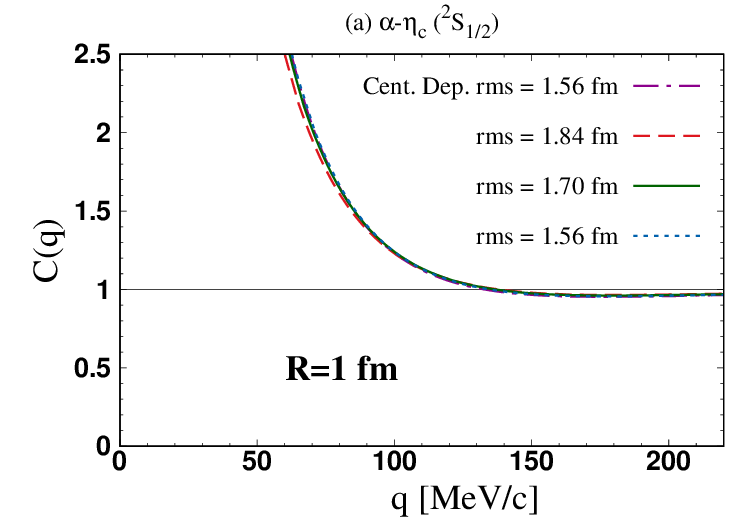}    
	
	\includegraphics[scale=0.64]{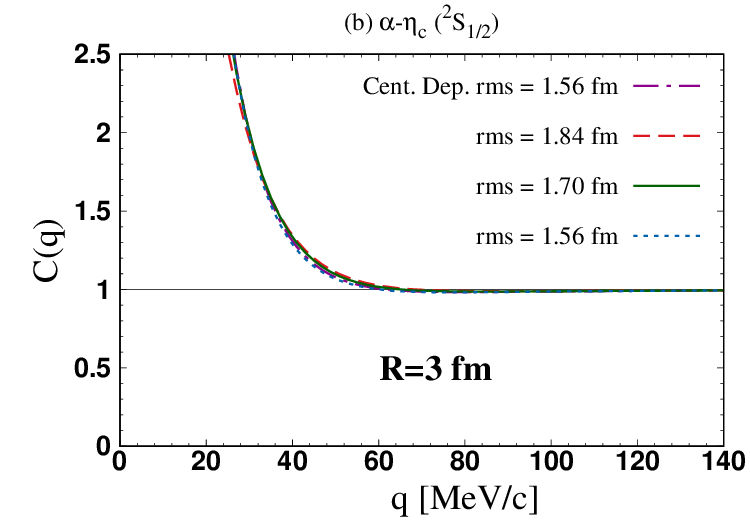} \includegraphics[scale=0.64]{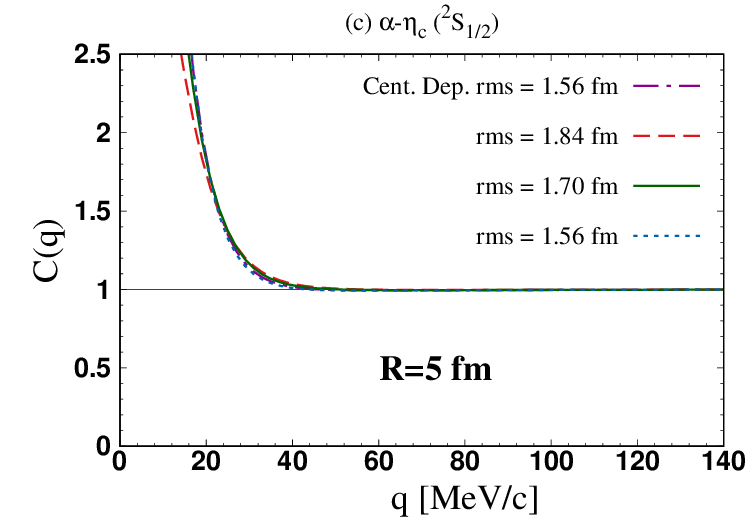}
	\caption{ The spin-$1/2$ $ \alpha\textrm{-} \eta_{c} $  correlation functions for three different source sizes: (a) $ R=1 $ fm, (b) $ R=3 $ fm and (c) $ R=5 $ fm. Symbols have the same description as in Fig.~\ref{fig:cq-kp-JPsi4S32Alpha}.
		\label{fig:cq-kp-EtacAlpha}   }
\end{figure*}

\begin{figure*}[hbt!]
	\centering
	\includegraphics[scale=0.64]{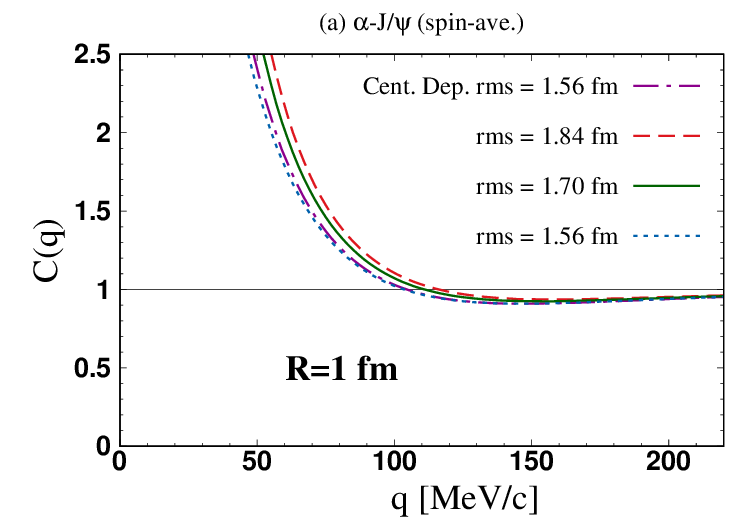}    
	
	\includegraphics[scale=0.64]{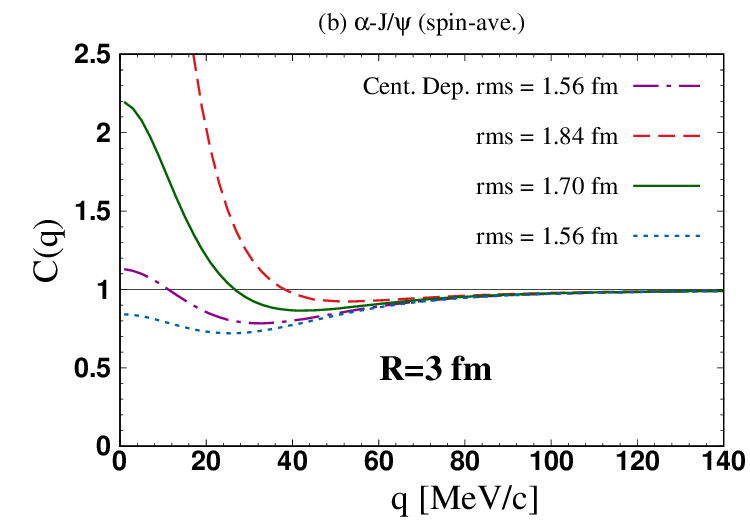} \includegraphics[scale=0.64]{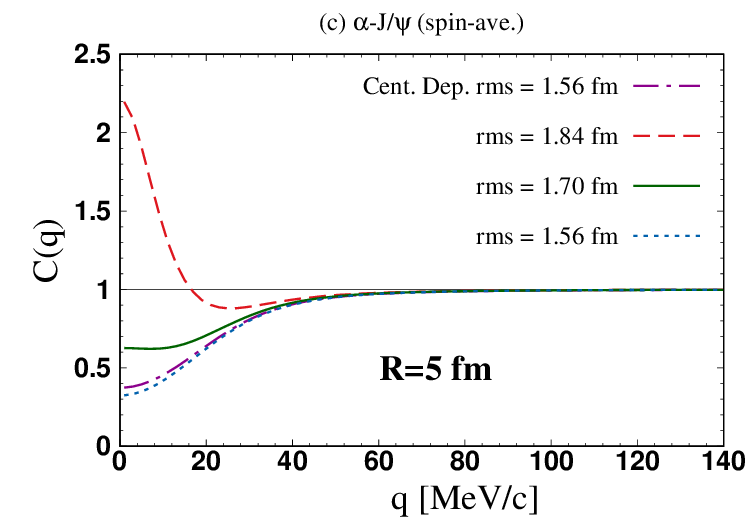}
	\caption{ The spin-averaged (spin-ave.) $ \alpha\textrm{-}J/\psi $  correlation functions for three different source sizes: (a) $ R=1 $ fm, (b) $ R=3 $ fm and (c) $ R=5 $ fm. Symbols have the same description as in Fig.~\ref{fig:cq-kp-JPsi4S32Alpha}.
		\label{fig:cq-kp-Jpsi-spin-aveAlpha-R}   }
\end{figure*}

Correlation functions for the $\alpha\textrm{-}c\bar{c}$ system are derived from the $U_{\alpha\textrm{-}c\bar{c}}$ potentials
using the KP formula~\eqref{eq:kp} for three different source sizes, $R=1,3 $ fm and $ 5 $ fm.
The results are shown by Figs.~\ref{fig:cq-kp-JPsi4S32Alpha},~\ref{fig:cq-kp-JPsi2S12Alpha},~\ref{fig:cq-kp-EtacAlpha} and ~\ref{fig:cq-kp-Jpsi-spin-aveAlpha-R} for spin-$3/2$ $ \alpha\textrm{-}J/\psi $, spin-$1/2$ $ \alpha\textrm{-}J/\psi $, 
spin-$1/2$ $ \alpha\textrm{-}\eta_{c} $ and the spin-averaged $ \alpha\textrm{-}J/\psi $ potentials, respectively.

 By altering the rms radius from 1.56 fm to 1.84 fm, the binding energy is decreased according to results in Table \ref{tab:Charm-alpha-para}. Conversely, enhancement of the corresponding correlation functions at small momenta is observed in Figs. \ref{fig:cq-kp-JPsi4S32Alpha} and \ref{fig:cq-kp-JPsi2S12Alpha}. It is found that the result obtained with a more attractive potential—for example, the $ \alpha\textrm{-}J/\psi $ potential with an rms radius of 1.54 fm—exhibits suppression or a bump structure depending on the source size \(R\). This behavior is representative of \(C(q)\) for an attractive interaction that supports a relatively loosely bound state. By contrast, correlation functions computed with a less attractive potential (for instance, the $ \alpha\textrm{-}J/\psi $ potential with an rms radius of 1.84 fm) display enhancement in the low-momentum region, characteristic of an attractive interaction without a bound state (or at least including a loosely bound state), though a dip structure is found in the intermediate-momentum region around \(q \sim 200\) MeV/c. Moreover, a very strong enhancement of the correlation at small momentum is indicative of a significantly large scattering length. 
Concisely, through a comparison of different models, it is found that stronger absorption of the potential leads to greater suppression of the correlation function.

The choice of these source sizes is motivated by values suggested in analyses of the $\Lambda \alpha$ correlation function~\cite{jinno2024femtoscopic}.
Since the charge radius of the $\alpha$-particle is \(1.70 \pm 0.14\) fm~\cite{PhysRevC.109.L012201}, a source radius of $R=1$ fm may appear small for $\alpha$-particle emission. However, it is discussed in Ref.~\cite{cats} that the term $4\pi r^{2}S_{r}(r)$ in Eq.~\eqref{eq:kp} describes the probability distribution of relative distances, where $S_{r}(r)$ is a Gaussian source function with a width of $\sqrt{2} R$, as shown in~\cite{cho2017exotic,kamiya2024}. Consequently, with $R=1$ fm, the average distance between emitted pairs is roughly $\left\langle r \right\rangle = 4 R / \sqrt{\pi} \approx 2.26$ fm, which is substantially larger than the value of $R$.

Divergence in the results for different potential parameters is observed in Figs.~\ref{fig:cq-kp-JPsi4S32Alpha},~\ref{fig:cq-kp-JPsi2S12Alpha}, and~\ref{fig:cq-kp-Jpsi-spin-aveAlpha-R} at low momentum ($q \lesssim 60$ MeV/c) for the spin-$3/2$ $\alpha\textrm{-}J/\psi$, spin-$1/2$ $\alpha\textrm{-}J/\psi$, and spin-averaged $\alpha\textrm{-}J/\psi$ systems, respectively. This divergence is particularly pronounced for the source with a size of $R = 3$ fm, as illustrated in panel (b) of each figure. 

On the other hand, this deviation is almost negligible in the results for spin-$1/2$ $ \alpha\textrm{-}\eta_{c} $ as shown by Fig.~\ref{fig:cq-kp-EtacAlpha} $R =1, 3$ and 5 fm. Remembering that the difference among the models mainly lies in their behavior at short range, therefore, in this case it seems, they have almost same behavior.

A comparison between two density functions of $ ^{4}\textrm{He}$—the CD density (Eq.~\eqref{eq:ctr_depRHO}) and the simple SG density (Eq.~\eqref{eq:gauss-dist})—with the same rms radius of 1.56 fm reveals notable differences. Despite nearly identical $\alpha\textrm{-}J/\psi$ binding energies for both models, their respective correlation functions exhibit distinct behaviors, as demonstrated in Figs.~\ref{fig:cq-kp-JPsi4S32Alpha},~\ref{fig:cq-kp-JPsi2S12Alpha}, and~\ref{fig:cq-kp-Jpsi-spin-aveAlpha-R}.
This result suggests that the future measurement of the $\alpha\textrm{-}J/\psi$ correlation function for$ R\geq3$ fm, source can constrain the $N\textrm{-}J/\psi$ interaction at high densities.
Specifically, the $\alpha\textrm{-}J/\psi$ correlation functions derived from the CD density model are consistently stronger than those obtained from the simple SG density distribution model.

In particular, the correlation function $C(q)$ with the spin-$1/2$ $\alpha\textrm{-}J/\psi$ potential exhibits a qualitatively different behavior from the others, which is attributed to the existence of a bound state associated with this potential.
As shown in Fig.~\ref{fig:v_CharmAlpha} and summarized in Table~\ref{tab:Charm-alpha-para}, the $NJ/\psi(^{2}S_{1/2})$ potential is more attractive than the others, leading to an enhancement in $C(q)$. 
However, this conclusion is not straightforwardly supported by Fig.~\ref{fig:CharmN}. 
The strong enhancement observed with the $\alpha\textrm{-}J/\psi(^{4}S_{3/2})$ and spin-averaged $\alpha\textrm{-}J/\psi$ potentials
at small $q$ reflects their relatively large scattering length, (see Table~\ref{tab:Charm-alpha-para}).

Consequently, different spin-dependent potentials can be distinguished through measurements of $C_{\alpha\textrm{-}c\bar{c}}(q)$,
especially for larger source sizes $R=3$ to $ 5 $ fm.
It is found that the $\alpha\textrm{-}J/\psi(^{2}S_{1/2})$ result shows either suppression or a bump depending on the source size, which is characteristic of attractive interactions with a bound state. 
Conversely, the correlation functions corresponding to $\alpha\textrm{-}J/\psi(^{4}S_{3/2})$ and the spin-averaged $\alpha\textrm{-}J/\psi$ display enhancements at low $q$, characteristic of attractive interactions without a bound state,
along with a dip at intermediate momentum around $ 60 $ MeV/c. The dip becomes less prominent when $R=1$ fm. This behavior suggests a simple attractive potential shape~\cite{Morita2016, kamiya2024}. While in the case of $\alpha\textrm{-}\eta_c(^{2}S_{1/2})$, 
a dip structure in the intermediate momentum region ($q\backsim200$ MeV/c) hardly is found, the dip structure is
more prominent in $ C\left(q\right) $ with a small source, $R = 1$ fm.

\begin{figure*}[hbt!]
	\centering
	\includegraphics[scale=0.64]{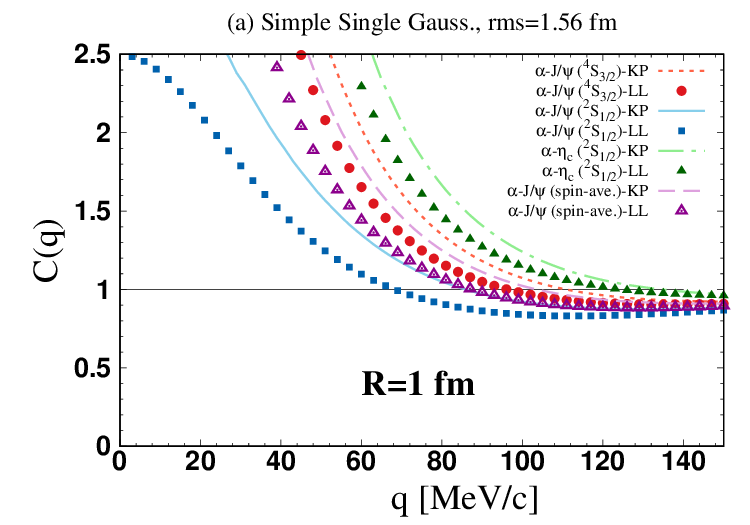}    
	
	\includegraphics[scale=0.64]{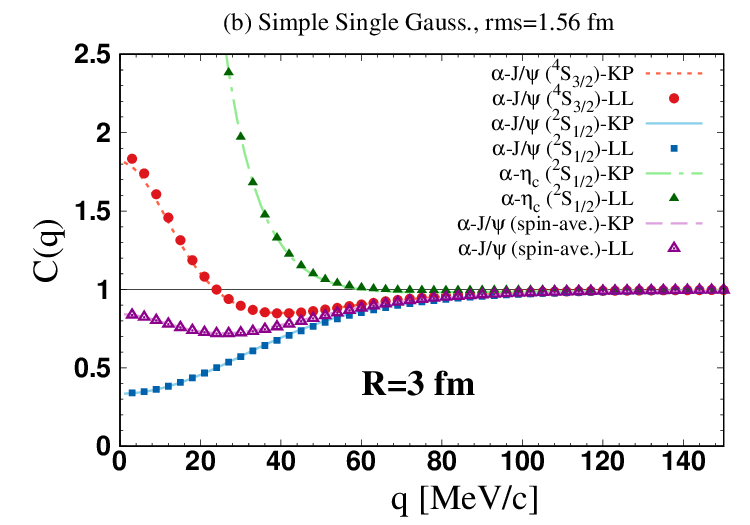} \includegraphics[scale=0.64]{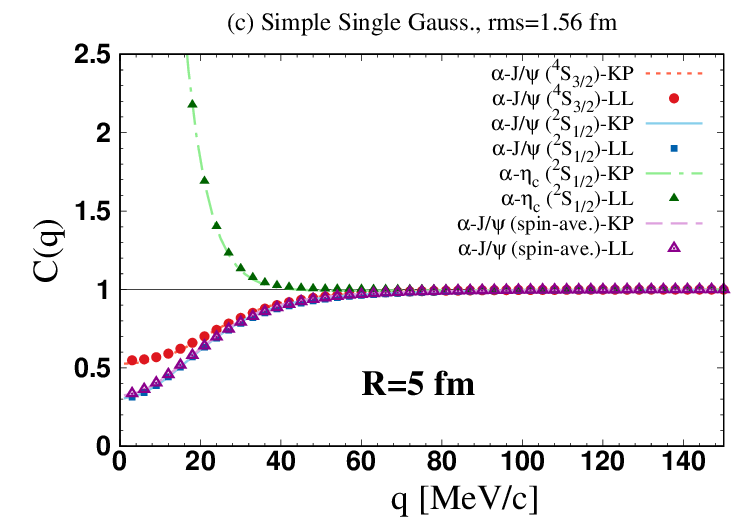}
	\caption{ 
		Employing the simple single Gaussian density distribution of $\alpha$ (Eq.~\eqref{eq:gauss-dist}), the $ \alpha\textrm{-}c\bar{c} $ correlation functions, $ C\left(q\right) $, are calculated. The results of the KP, Eq.~\eqref{eq:kp} and the LL, Eq.~\eqref{eq:ll} formulae are compared for three different source sizes (a) $ R=1$, (b) $ R=3$ and (c) $ R=5$ fm. 
		The results (from KP and LL formulae) are shown for spin-$3/2$ $ \alpha\textrm{-}J/\psi $ (red dotted line and full  circle), 
		spin-$1/2$ $ \alpha\textrm{-}J/\psi$ (blue solid line and full square), spin-$ 1/2 $ $\alpha\textrm{-}\eta_{c}$ (green dashed line and full  triangle) 
		and the spin-averaged $ \alpha\textrm{-}J/\psi $ (dash-dotted purple line and open triangle) models of potential. 
		\label{fig:charmAlpha-SG-156_cq_kp_ll}}
\end{figure*}

\begin{figure*}[hbt!]
	\centering
	\includegraphics[scale=0.64]{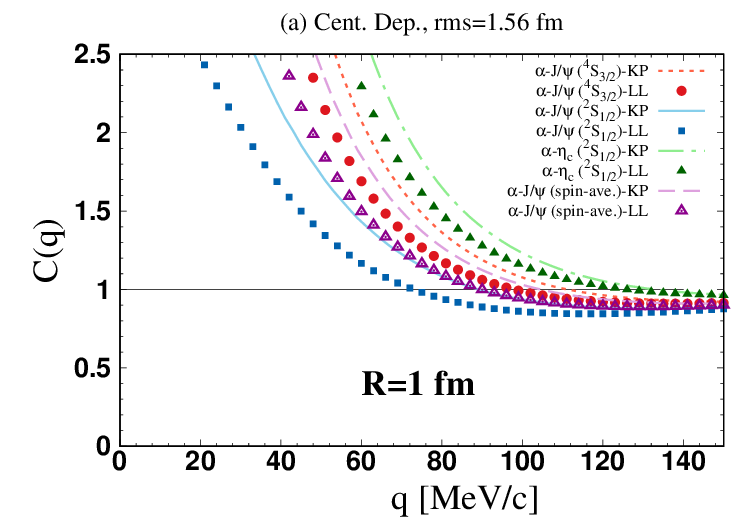}    
	
	\includegraphics[scale=0.64]{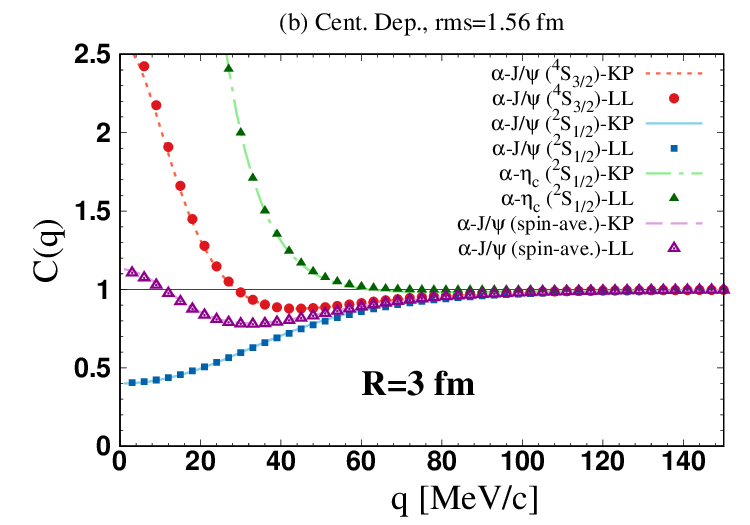} \includegraphics[scale=0.64]{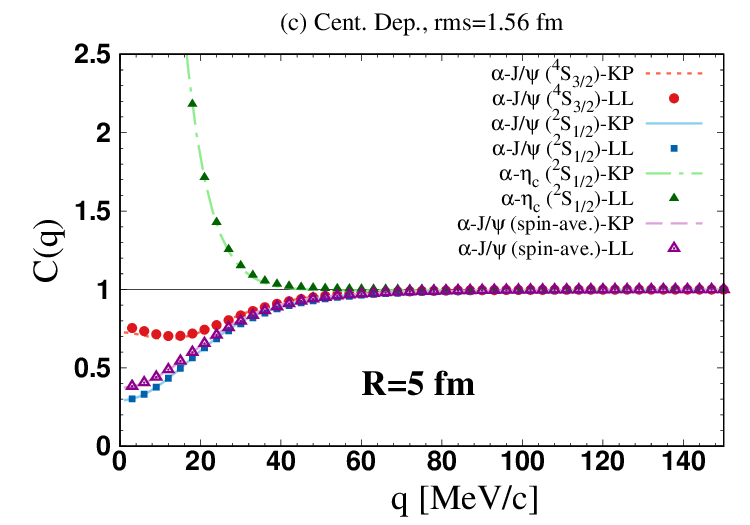}
	\caption{ 
		Employing the central depression density distribution of $\alpha$ (Eq.~\eqref{eq:ctr_depRHO}), the $ \alpha\textrm{-}c\bar{c} $ correlation functions  are calculated. The results of the  KP, Eq.~\eqref{eq:kp} and the LL, Eq.~\eqref{eq:ll} formulae are compared for three different source sizes (a) $ R=1$, (b) $ R=3$ and (c) $ R=5$ fm. 
		The results (from KP and LL formulae) are shown for spin-$3/2$ $ \alpha\textrm{-}J/\psi $ (red dotted line and full  circle), 
		spin-$1/2$ $ \alpha\textrm{-}J/\psi$ (blue solid line and full  square), spin-$ 1/2 $ $\alpha\textrm{-}\eta_{c}$ (green dashed line and full  triangle) 
		and the spin-averaged $ \alpha\textrm{-}J/\psi $ (dash-dotted purple line and open triangle) models of potential. 
		\label{fig:charmAlpha-CD-156_cq_kp_ll}}
\end{figure*}

The $\alpha\textrm{-}c\bar{c}$ correlation function is calculated from the scattering length and the effective range listed in Table~\ref{tab:Charm-alpha-para} using the LL formula~\eqref{eq:ll}, and the results are compared with those obtained from the KP formula in Fig.~\ref{fig:charmAlpha-SG-156_cq_kp_ll} by employing the simple single Gaussian density distribution of $\alpha$ (Eq.~\eqref{eq:gauss-dist}) and in Fig.~\ref{fig:charmAlpha-CD-156_cq_kp_ll} by using  the central depression density distribution of $\alpha$ (Eq.~\eqref{eq:ctr_depRHO}),  for the three source sizes.
It is shown in Figs.~\ref{fig:charmAlpha-SG-156_cq_kp_ll} and Fig.~\ref{fig:charmAlpha-CD-156_cq_kp_ll} that, for $R=1$ fm, the LL approach yields significantly different results from the KP formula at low momentum. This discrepancy arises because the LL formula tends to be an inadequate approximation when the source size is smaller than the range of the interaction, which, in interactions involving nuclei, is typically $\gtrsim 3$ fm~\cite{jinno2024femtoscopic}. As can be seen from Fig.~\ref{fig:v_CharmAlpha}, the derived potentials of $\alpha\textrm{-}c\bar{c}$ well become zero at a distance of about $3$ fm.
Conversely, for larger source sizes ($R \geq 3$ fm), the LL approximation converges with the KP results, indicating its validity in those regimes.

\begin{figure*}[hbt!]
	\centering
	\includegraphics[scale=0.64]{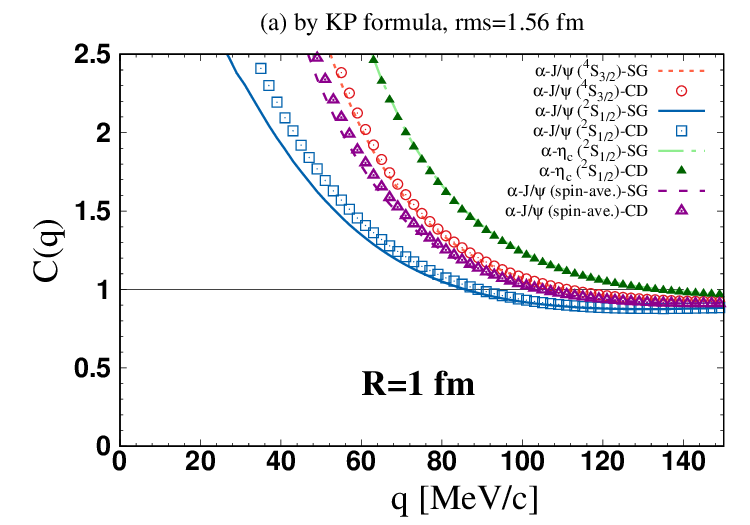}    
	
	\includegraphics[scale=0.64]{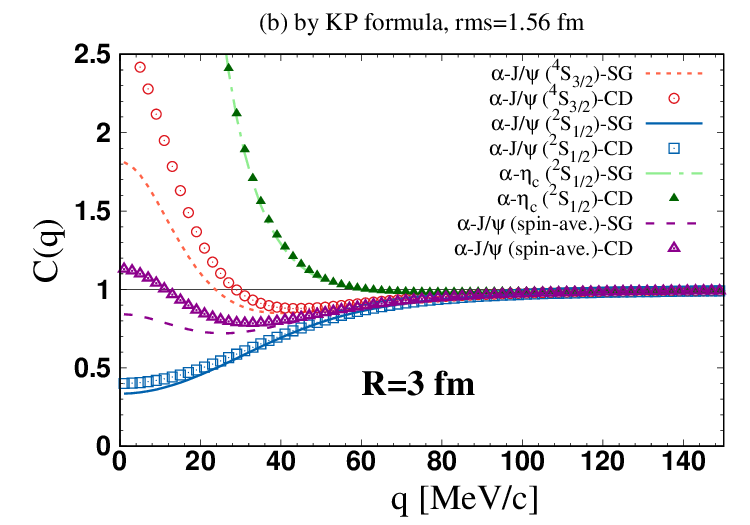} \includegraphics[scale=0.64]{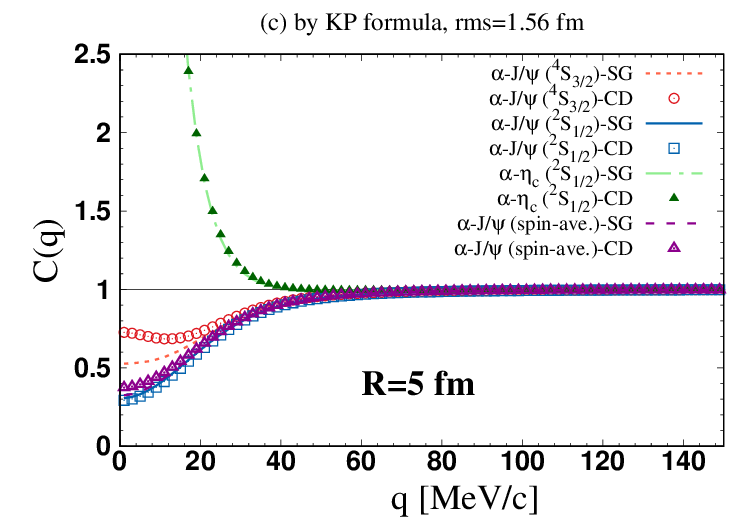}
	\caption{ 
		The obtained $ \alpha\textrm{-}c\bar{c} $ correlation functions from  the simple single Gaussian (SG),~Eq.~\eqref{eq:gauss-dist} and the central depression (CD), Eq.~\eqref{eq:ctr_depRHO} density distribution of $\alpha$, via KP formula, Eq.~\eqref{eq:kp} are compared for three different source sizes (a) $ R=1$, (b) $ R=3$ and (c) $ R=5$ fm. 
		The results (from SG and CD distribution) are shown for spin-$3/2$ $ \alpha\textrm{-}J/\psi $ (red dotted line and open circle), 
		spin-$1/2$ $ \alpha\textrm{-}J/\psi$ (blue solid line and open square), spin-$ 1/2 $ $\alpha\textrm{-}\eta_{c}$ (green dash-dotted line and full  triangle) 
		and the spin-averaged $ \alpha\textrm{-}J/\psi $ (dash purple line and open triangle) models of potential.  
		\label{fig:charmAlpha-SGCD-156_cq_kp_kp}}
\end{figure*}

\begin{figure*}[hbt!]
	\centering
	\includegraphics[scale=0.64]{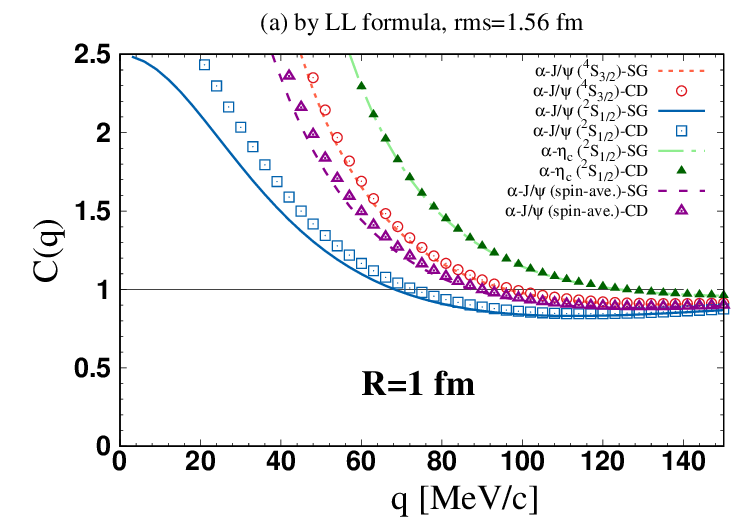}    
	
	\includegraphics[scale=0.64]{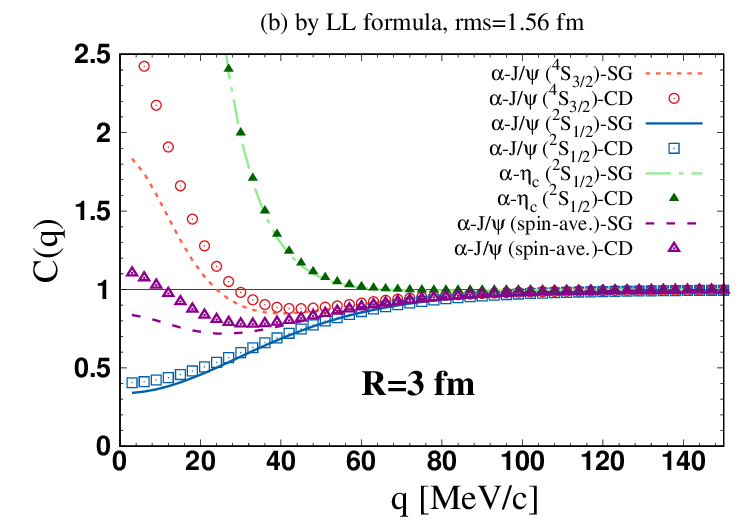} \includegraphics[scale=0.64]{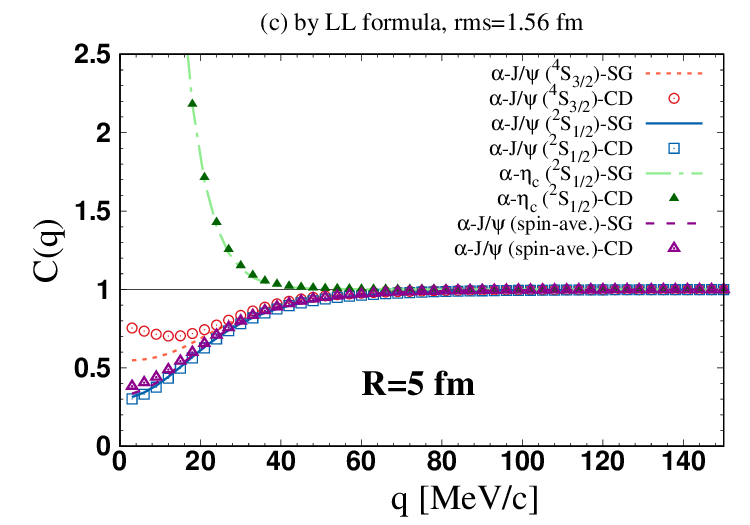}
	\caption{ 
		The obtained $ \alpha\textrm{-}c\bar{c} $ correlation functions from  the simple single Gaussian (SG),~Eq.~\eqref{eq:gauss-dist} and the central depression (CD), Eq.~\eqref{eq:ctr_depRHO} density distribution of $\alpha$, via LL formula, Eq.~\eqref{eq:ll} are compared for three different source sizes (a) $ R=1$, (b) $ R=3$ and (c) $ R=5$ fm. 
		The results (from SG and CD distribution) are shown for spin-$3/2$ $ \alpha\textrm{-}J/\psi $ (red dotted line and open circle), 
		spin-$1/2$ $ \alpha\textrm{-}J/\psi$ (blue solid line and open square), spin-$ 1/2 $ $\alpha\textrm{-}\eta_{c}$ (green dash-dotted line and full  triangle) 
		and the spin-averaged $ \alpha\textrm{-}J/\psi $ (dash purple line and open triangle) models of potential. 
		\label{fig:charmAlpha-SGCD-156_cq_ll_ll}}
\end{figure*}

To assess the role of the nuclear matter distribution function of $ ^{4}\textrm{He}$ in the $ \alpha\textrm{-}c\bar{c} $ correlation functions, which share the same rms radius of 1.56 fm, two comparisons were conducted. These comparisons are illustrated in Fig.~\ref{fig:charmAlpha-SGCD-156_cq_kp_kp} and Fig.~\ref{fig:charmAlpha-SGCD-156_cq_ll_ll}.  

In the first comparison, the $ \alpha\textrm{-}c\bar{c} $ correlation functions derived from the simple SG and CD density distributions of the alpha particle were evaluated using the KP formula (Eq.~\eqref{eq:kp}), as shown in Fig.~\ref{fig:charmAlpha-SGCD-156_cq_kp_kp}. The second comparison involved the same correlation functions but was computed using the LL formula (Eq.~\eqref{eq:ll}), as depicted in Fig.~\ref{fig:charmAlpha-SGCD-156_cq_ll_ll}. Both comparisons were performed for three distinct source sizes: (a) $ R=1$ fm, (b) $ R=3$ fm, and (c) $ R=5$ fm.  

From the analysis of Figs.~\ref{fig:charmAlpha-SGCD-156_cq_kp_kp} and \ref{fig:charmAlpha-SGCD-156_cq_ll_ll}, it was observed that for the spin-$3/2$ $ \alpha\textrm{-}J/\psi$ system and the spin-averaged $ \alpha\textrm{-}J/\psi$ system, a notable discrepancy exists between the correlation functions calculated using the SG and CD distributions. This discrepancy is particularly pronounced for a source size of 3 fm. Such a difference could provide valuable insights into the nuclear matter distribution function within the alpha particle, thereby enhancing our understanding of its structural properties.

A concern exists that if charmonium and alpha particles are produced at different stages of a collision under varying thermodynamic conditions, the correlation signal between them will be extremely weak. Charmonium is generated early in the collision through hard scattering processes and is sensitive to the QGP, requiring high collision energies due to its large mass; otherwise, its production yield is very low. Conversely, alpha particles form later during the hadronic phase via nucleon coalescence and are predominantly produced at lower energies where the system is baryon-rich. The thermal model predicts a significant suppression of alpha particle production at high energies, which is why lower-energy collisions—such as those at the CERN-SPS (Super Proton Synchrotron) or GSI-FAIR (Facility for Antiproton and Ion Research)—are preferred. However, these low energies also substantially suppress charmonium production, according to the same model. While large datasets can provide overall charmonium yields, femtoscopy analyses require both particles to be produced simultaneously in the same event, making the expected correlation signal exceedingly small.

To tackle these challenges, it has been suggested that the quantities of light nuclei remain roughly constant during the transition from chemical to thermal freeze-out in the evolving hadron fireball, due to their ongoing formation and dissociation during this phase~\cite{VOVCHENKO2020135131}. However, challenges exist beyond the fact that the size of light nuclei is comparable to the spacing between hadrons in the fireball. For example, the formation time of a deuteron—around 100 fm/c, based on its inverse binding energy—is much longer than the typical timescale assumed for repeated formation and dissociation processes within the fireball~\cite{bazak2020production,PhysRevX.14.031051}.  Consequently, assuming continuous formation and dissociation within such a brief interval remains problematic. To address these issues, proponents of the thermal model propose that the final nuclei originate from compact, colorless quark and gluon states present inside the fireball~\cite{andronic2018decoding}.

On the other hand,  in the kinetic approach, alpha particles may also be produced at the early stages of collisions and may survive subsequent hadronic scatterings due to factors such as quantum Mott effects and the rapid expansion of the fireball.  
The yields of light nuclei in central Au+Au collisions at incident energies from $0.25A$ GeV to $1.0A$ GeV have been reasonably reproduced by Wang et al. in Ref.~\cite{PhysRevC.108.L031601}, in comparison with measurements by the FOPI Collaboration~\cite{REISDORF2010366}.  
Recently, the yield of $ J/\psi $ mesons, produced in proton–nucleus and nucleus–nucleus collisions, has been estimated within a Glauber-model framework for upcoming low-energy heavy-ion experiments at SPS and FAIR~\cite{CHATTERJEE2023122554}.  
It is noted that an actual challenge may arise from the very low multiplicities of both alpha particles and $ J/\psi $ at FAIR/NICA energies, which could complicate the experimental measurement of alpha–charmonium correlations.

Overall, direct experimental evidence elucidating the microscopic mechanisms of nucleus formation remains elusive. Femtoscopy offers a complementary approach by examining correlations between momenta of light nuclei and other particles, providing direct insight into the underlying microscopic processes responsible for light nuclei formation. This technique has been effectively employed by the ALICE Collaboration to analyze various hadron pair correlations produced in pp and p–Pb collisions at the LHC~\cite{PhysRevX.14.031051} and related references, shedding light on their residual strong interactions.

\section{Summary and conclusions} \label{sec:Summary-and-conclusions}

Two-body  $ ^{4}\textrm{He}\left(\alpha\right)$-charmonium $ \left(c\bar{c}\right) $ potentials in 
the SFP approach were built by using a realistic low-energy $ NJ/\psi$ and $ N\eta_{c} $ interactions based on $\left(2+1\right)$-flavor configurations with nearly physical pion mass $ m_{\pi}= 146 $ MeV~\cite{yan2022prd}. 
The potentials of alpha-charmonium are evaluated using various well-established matter distributions and the rms radius of $\textrm{\ensuremath{^{4}}He}$. 
Then, the obtained $ \alpha\textrm{-}c\bar{c} $ potentials fitted by a Woods-Saxon type function and employed as the input to solve 
the Schr\"{o}dinger equation.

Numerical analysis revealed that the $ \alpha\textrm{-}J/\psi $ system appears to be loosely bound, with the central binding energy estimated to lie within the range of 0.1–0.6 MeV. In contrast, no bound or resonance state (relative to the $ \alpha \textrm{-} c\bar{c} $ threshold) was identified for the spin-$ 1/2 $ $\alpha\textrm{-}\eta_{c}$ system.

I applied femtoscopy technique to predict $ \alpha\textrm{-}c\bar{c} $ momentum correlation functions in
high-energy nuclear collisions to look for an additional and alternative source of knowledge relevant to the
$ N\textrm{-}c\bar{c} $ interaction.  
Employing the derived $ \alpha\textrm{-}c\bar{c} $ potentials, correlation functions were calculated  using the KP formula for three different source sizes, $ R=1,3 $ fm and $ 5 $ fm.     
Furthermore, correlation functions were examined within the LL approximation and compared with the results of using the KP formula. 
It is found that the LL approximates different results at small source sizes,
this indicates that the $ \alpha\textrm{-}c\bar{c} $ has relatively long-range interaction.

In this exploratory study, the selection of source sizes $R = 1$, $3$, and $5$ fm was based on previous investigations of the two-hadron correlation function in $pp$ collisions and heavy ion collisions~\cite{jinno2024femtoscopic,kamiya2024}. 
The validity of the KP formula, Eq.~\eqref{eq:kp}, is upheld when the two correlated particles are regarded
as well-separated, point-like particles.
For composite particles such as the $\alpha$ particle, since the possibility of simultaneous formation exists, the effective source size is required to be larger than that for the emission of any single hadron~\cite{mrowczynski2019hadron,bazak2020production,StanislawPRC2021}.
Consequently, a $ 5 $-body problem involving two protons, two neutrons, and a $c\bar{c}$ pair is essentially encountered, with the formation of the alpha particle and its correlation with the $c\bar{c}$ occurring simultaneously.
The consideration of this effect will be addressed in future work. 

The numerical analysis showed that the differences in spin-dependent $ N\textrm{-}c\bar{c} $ interactions, i.e, spin-$3/2$ $ NJ/\psi $, spin-$1/2$ $ NJ/\psi$, spin-$ 1/2 $ $N\eta_{c}$ and the spin-averaged $ NJ/\psi $ interactions, lead to well detectable differences in the $ \alpha\textrm{-}c\bar{c} $ correlation function, in particular by those from $ R \sim 3 $ fm source.

 It is noteworthy that in the present work the $ \alpha\textrm{-}J/\psi $ potentials in the spin channels $ ^{4}S_{3/2} $, $^{2}S_{1/2} $, as well as the spin-averaged $ \alpha\textrm{-}J/\psi $ potential, are obtained by substituting the corresponding $N\textrm{-}J/\psi $ potentials
$ \left(V_{N\textrm{-}c\bar{c}}\right)$ into  Eq.~\eqref{eq:V_alfaOmega} respectively.
As mentioned, this is a simple approximation; experimentally, the alpha$ \textrm{-}J/\psi $ interaction is expected to be spin-averaged (with the spin-dependent nucleon$ \textrm{-}J/\psi $ interactions) because the four nucleons in alpha/(NNNN) are not spin-polarized.

In reality, the formation of charmonium such as $ J/\psi$ through four-nucleon coalescence involves a complex five-body scattering process comprising four nucleons and a charmonium.  A full-scale, ab initio, or highly accurate calculation of this five-body problem—such as with the no-core shell model~\cite{7q75-9bbx} or by solving the Faddeev–Yakubovsky equations~\cite{10.3389/fphy.2019.00251}—would be required for an exact description.
However, such comprehensive calculations are beyond the scope of this work and are left as an avenue for future research.

Additionally, a comparison was performed between two density functions of $ ^{4}\textrm{He}$—the central depression (CD) and the simple single Gaussian (SG) density—both of which exhibit an identical rms radius of 1.56 fm. Significant discrepancies in the results were observed. While the binding energies of the $\alpha\textrm{-}J/\psi$ system for the two models were found to be nearly identical, their respective correlation functions displayed markedly distinct behaviors. This discrepancy may provide valuable insights into the nuclear matter distribution function of the alpha particle, thereby enhancing the understanding of its structural properties.

In heavy-ion collisions, light nuclei such as alpha particles experience medium-induced modifications to their structure~\cite{lombardo2023clusters,ccbm-lk5v}. Key mechanisms include Pauli blocking by surrounding nucleons, self-energy shifts from mean fields, and screening of interactions, all of which can lower the in-medium binding energy and push the alpha cluster toward dissolution at a density (the Mott point) that depends on temperature and isospin. 
Finite temperature further populates excited states and broadens energy levels, altering the effective size and density distribution of the alpha cluster. Consequently, the alpha’s r.m.s. radius and surface diffuseness can change, and the formation probability of alphas becomes sensitive to local density and temperature, with potential observable consequences in yields, spectra, and correlations~\cite{PhysRevLett.108.062702}.

To study these effects, researchers use frameworks such as quantum-statistical approaches~\cite{PhysRevC.84.055804}, cluster-virial or density-dependent relativistic mean-field  models~\cite{PhysRevC.81.015803}, few-body in-medium calculations~\cite{ccbm-lk5v}, and transport simulations that incorporate cluster degrees of freedom. Observable consequences include modifications to alpha yields and chemical equilibration, altered momentum spectra and collective flow, and changes in two-particle correlations and inferred radii. Open questions remain about accurately determining the Mott density under realistic, evolving collision conditions, disentangling in-medium effects from production mechanisms, and achieving a consistent description across equations of state and transport dynamics.

In experiments aimed at measuring the $ \alpha\textrm{-}c\bar{c} $ correlation, the use of high-energy collisions with a central energy $ \sqrt{s_{NN}} < 10 $ is favored, as the production of $\alpha$ particles is estimated to occur under these conditions~\cite{ANDRONIC2011203}.
For this reason, it is expected that the $ \alpha\textrm{-}c\bar{c} $ correlation will be experimentally accessible at facilities such as FAIR~\cite{cbm}, NICA (Nuclotron-based Ion Collider fAcility)~\cite{KEKELIDZE2013945c}, HIAF (High Intensity Heavy-ion Accelerator Facility)~\cite{HIAF} and J-PARC HI~\cite{j-park}, in order to clarify the nature of the $ \alpha\textrm{-}c\bar{c} $ interaction and its relationship to the $ N\textrm{-}c\bar{c} $ interactions.

\section*{Author contributions}
The author confirms the sole responsibility for the conception of the study, presented results and
manuscript preparation.

\section*{Competing interests}
The author(s) declare no competing interests.

\section*{Funding Declaration}
The author(s) received no financial support for the research, authorship, and/or publication of this article.

\section*{Data availability}
The datasets used and/or analyzed in this study are available at~\url{https://doi.org/10.5281/zenodo.19532848} and can also be obtained from the corresponding author upon reasonable request.

\bibliography{Refs.bib}
\end{document}